\documentclass[iop,apj]{emulateapj}

\pdfoutput=1

\slugcomment{Nov 30, 2015}

\shorttitle{The Chandra COSMOS Legacy Survey}
\shortauthors{Civano et al.}

\def\jcap{Journal of Cosmology and Astroparticle Physics}

\def\chandra{{\it Chandra\/}}
\def\xmm{{XMM-{\it Newton\/}}}
\def\leg{{\em COSMOS-Legacy}\/}

\def\arcsec{$^{\prime\prime}$}
\def\amin{$^{\prime}$}
\newcommand{\cgs}{ ${\rm erg~cm}^{-2}~{\rm s}^{-1}$} 
\newcommand{\lum}{\rm erg~s$^{-1}$}

\begin{document}


\title{The \chandra\ COSMOS Legacy survey: overview and point source catalog}


\author{F. Civano\altaffilmark{1,2}, S. Marchesi\altaffilmark{1,3,2}, A. Comastri\altaffilmark{4}, M.C. Urry\altaffilmark{1}, M. Elvis\altaffilmark{2}, N. Cappelluti\altaffilmark{4}, S. Puccetti\altaffilmark{5}, M. Brusa\altaffilmark{3,4},  G. Zamorani\altaffilmark{4}, G. Hasinger\altaffilmark{6}, T. Aldcroft\altaffilmark{2}, D. M. Alexander\altaffilmark{7}, V. Allevato\altaffilmark{8}, H. Brunner\altaffilmark{9}, P. Capak\altaffilmark{10,11}, A. Finoguenov\altaffilmark{8}, F. Fiore\altaffilmark{12}, A. Fruscione\altaffilmark{2}, R. Gilli\altaffilmark{4}, K. Glotfelty\altaffilmark{2}, R. E. Griffiths\altaffilmark{13}, H. Hao\altaffilmark{2}, F. A. Harrison\altaffilmark{14}, K. Jahnke\altaffilmark{15}, J. Kartaltepe\altaffilmark{16,17}, A. Karim\altaffilmark{18}, S. M. LaMassa\altaffilmark{1}, G. Lanzuisi\altaffilmark{3,4}, T. Miyaji\altaffilmark{19,20}, P. Ranalli\altaffilmark{21}, M. Salvato\altaffilmark{9}, M. Sargent\altaffilmark{22}, N. J. Scoville\altaffilmark{11}, K. Schawinski\altaffilmark{23}, E. Schinnerer\altaffilmark{15,24}, J. Silverman\altaffilmark{25}, V. Smolcic\altaffilmark{26}, D. Stern\altaffilmark{27}, S. Toft\altaffilmark{28}, B. Trakhenbrot\altaffilmark{23}, E. Treister\altaffilmark{29}, C. Vignali\altaffilmark{3,4}}

\altaffiltext{1}{Yale Center for Astronomy and Astrophysics, 260 Whitney Avenue, New Haven, CT 06520, USA}
\altaffiltext{2}{Harvard-Smithsonian Center for Astrophysics, 60 Garden Street, Cambridge, MA 02138, USA}
\altaffiltext{3}{Dipartimento di Fisica e Astronomia, Universit\`a di Bologna, viale Berti Pichat 6/2, 40127 Bologna, Italy}
\altaffiltext{4}{INAF--Osservatorio Astronomico di Bologna, via Ranzani 1, 40127 Bologna, Italy}
\altaffiltext{5}{ASDC--ASI, Via del Politecnico, 00133 Roma, Italy}
\altaffiltext{6}{Institute for Astronomy, 2680 Woodlawn Drive, University of Hawaii, Honolulu, HI 96822, USA}
\altaffiltext{7}{Centre for Extragalactic Astronomy, Department of Physics, Durham University, South Road, Durham DH1 3LE, UK}
\altaffiltext{8}{Department of Physics, University of Helsinki, Gustaf H\"allstr\"omin katu 2a, FI-00014 Helsinki, Finland}
\altaffiltext{9}{Max-Planck-Institut f{\"u}r extraterrestrische Physik, Giessenbachstrasse 1, D-85748 Garching bei M{\"u}nchen, Germany}
\altaffiltext{10}{Infrared Processing and Analysis Center (IPAC), 1200 East California Boulevard, Pasadena, California 91125, USA} 
\altaffiltext{11}{California Institute of Technology, 1200 East California Boulevard, Pasadena, California 91125, USA}
\altaffiltext{12}{INAF - Osservatorio Astronomico di Roma, via Frascati 33, Monte Porzio Catone, I00040, Italy}
\altaffiltext{13}{Physics \& Astronomy Dept., Natural Sciences Division, University of Hawaii at Hilo, 200 W. Kawili St., Hilo, HI 96720, USA}
\altaffiltext{14}{Cahill Center for Astronomy and Astrophysics, California Institute of Technology, 1216 E. California Blvd, Pasadena, CA, 91125 USA}
\altaffiltext{15}{Max Planck Institute for Astronomy, Koenigstuhl 17, D-69117 Heidelberg, Germany}
\altaffiltext{16}{National Optical Astronomy Observatory, 950 N. Cherry Ave, Tucson AZ 85719, USA}
\altaffiltext{17}{School of Physics and Astronomy, Rochester Institute of Technology, 84 Lomb Memorial Dr., Rochester, NY 14623, USA}
\altaffiltext{18}{Argelander-Institut f\"ur Astronomie, Universit\"at Bonn, Auf dem H\"ugel 71, D-53121 Bonn, Germany}
\altaffiltext{19}{Instituto de Astronom\'ia sede Ensenada, Universidad Nacional Aut\'onoma de M\'exico, Km. 103, Carret. Tijunana-Ensenada, Ensenada, BC, Mexico}
\altaffiltext{20}{University of California San Diego, Center for Astrophysics and Space Sciences, 9500 Gilman Drive, La Jolla, CA 92093-0424, USA}
\altaffiltext{21}{AASARS, National Observatory of Athens, 15236 Penteli, Greece}
\altaffiltext{22}{Astronomy Centre, Department of Physics and Astronomy, University of Sussex, Brighton, BN1 9QH, UK}
\altaffiltext{23}{Institute for Astronomy, Department of Physics, ETH Zurich, Wolfgang-Pauli-Strasse 27, CH-8093 Zurich, Switzerland}
\altaffiltext{24}{National Radio Astronomy Observatory, Pete V. Domenici Science Operations Center, 1003 Lopezville Road, Socorro, NM 87801, USA}
\altaffiltext{25}{Kavli Institute for the Physics and Mathematics of the Universe (Kavli IPMU, WPI), Todai Institutes for Advanced Study, The University of Tokyo, Kashiwa 277-8583, Japan}
\altaffiltext{26}{Department of Physics, University of Zagreb, Bijeni\v{c}ka cesta 32, HR-10000 Zagreb, Croatia}
\altaffiltext{27}{Jet Propulsion Laboratory, California Institute of Technology, 4800 Oak Grove Drive, Pasadena, CA 91109, USA}
\altaffiltext{28}{Dark Cosmology Centre, Niels Bohr Institute, University of Copenhagen, Juliane Mariesvej 30, DK-2100 Copenhagen, Denmark}
\altaffiltext{29}{Universidad de Concepci\'{o}n, Departamento de Astronom\'{\i}a, Casilla 160-C, Concepci\'{o}n, Chile}


\begin{abstract}
The \leg\ survey is a 4.6 Ms \chandra\ program that has imaged 2.2 deg$^2$ of the COSMOS field with an effective exposure of $\simeq$160 ks over the central 1.5 deg$^2$ and of $\simeq$80 ks in the remaining area. The survey is the combination of 56 new observations, obtained as an X-ray Visionary Project, with the previous C-COSMOS survey. We describe the reduction and analysis of the new observations and the properties of 2273 point sources detected above a spurious probability of 2$\times 10^{-5}$. We also present the updated properties of the C-COSMOS sources detected in the new data. The whole survey includes 4016 point sources (3814, 2920 and 2440 in the full, soft and hard band). The limiting depths are 2.2 $\times$ 10$^{-16}$, 1.5 $\times$ 10$^{-15}$ and 8.9$\times$ 10$^{-16}$ \cgs\ in the 0.5-2, 2-10 and 0.5-10 keV bands, respectively. The observed fraction of obscured AGN with column density $> 10^{22}$ cm$^{-2}$ from the hardness ratio (HR) is $\sim$50$^{+17}_{-16}$\%. Given the large sample, we compute source number counts in the hard and soft bands, significantly reducing the uncertainties of 5-10\%. For the first time, we compute number counts for obscured (HR$>$-0.2) and unobscured (HR$<$-0.2) sources and find significant differences between the two populations in the soft band. Due to the un-precedent large exposure, \leg\ area is 3 times larger than surveys at similar depth and its depth is 3 times fainter than surveys covering similar area. The area--flux region occupied by \leg\ is likely to remain unsurpassed for years to come. 
\end{abstract}



\section{Introduction}

One of the most active but least known epochs in astrophysics is the period between re-ionization (z$\geq$8, i.e. when the Universe was less than $\sim$0.6 Gyr old), 
where the growth of structures becomes highly non-linear and the first stars form, and z$\sim$2 ($\sim$3.25 Gyr old), 
where major virialization occurs and star formation (SF) and supermassive black hole (SMBH) accretion peak.

At these early times, the pre-cursors of the clusters and groups seen at $z\lesssim1$ have low density and are 
much larger, on both physical (Mpc) and observed (arcminutes) scales. Surveys for these large scale structures become
rapidly more efficient as the dimension of the survey exceeds the structure's typical sizes ($\sim$15\amin). 
Large area surveys (several times 15\amin\ wide) are essential for the
detection of these structures, which cannot be seen in smaller area surveys, however deep.

The equatorial 2 deg$^2$ COSMOS area (Scoville et al. 2007a) is the deepest, most complete survey accessible 
to both hemispheres (notably by both ALMA and the Karl G. Jansky VLA) and large enough to find high-redshift clusters. A significant investment of
640 {\em HST} orbits (Scoville et al. 2007b, Koekemoer et al. 2007), $620$h of {\em Spitzer} (Capak et al. in prep.), $260$h of {\em Herschel} (Lutz et al. 2011), 750 hrs of JVLA (Schinnerer et al. 2004, 2007, 2010 and Smolcic et al. 2014 and in prep.), over 300 nights of large ground-based telescopes VLT, Keck, Subaru, VISTA for both imaging and spectroscopy (Lilly et al. 2009, Hasinger et al. in prep., Taniguchi et al. 2007, McCraken et al. 2014) has been made in this field. 

The first homogeneous coverage in the X-rays of the whole COSMOS field was obtained with the \xmm\ satellite 
(1.5 Megaseconds; Hasinger et al. 2007, Cappelluti et al. 2009, Brusa et al. 2010). These observations have been crucial for characterizing the most 
luminous Active Galactic Nuclei (AGN) in COSMOS (e.g. Brusa et al. 2010, Mainieri et al. 2011, Allevato et al. 2011, Lusso et al. 2012, among others). 
The obscured AGN population of the COSMOS field can be studied by jointly using the XMM-COSMOS data with the $\sim$3 Megaseconds of {\it NuSTAR} (Harrison et al. 2012) time available, 
which led to the discovery of a Compton-thick AGN (with obscuration exceeding 10$^{24}$ cm$^{-2}$ equivalent hydrogen column density) in the field (Civano et al. 2015), that was not recognized as such by \xmm\ alone. 
Instead, for the faint and the high-z AGN population, that could be responsible for the re-ionization of the Universe (see Giallongo et al. 2015), 
\chandra\ (Weisskopf et al. 2002) is the preferred instrument. Indeed, the large (1.8 Ms) \chandra\ COSMOS survey 
(C-COSMOS; Elvis et al. 2009, E09; Puccetti et al. 2009, P09; Civano et al. 2012) 
has already contributed significantly to the study of the early epochs of the Universe, finding: 
three luminous AGN residing in protoclusters between z$\sim$4.55 and 5.3 (Capak et al. 2011); 
the largest sample of X-ray selected z$>$3 quasars in a contiguous field (81 sources, Civano et al. 2011); a precocious SMBH in a normal size galaxy at z$>$3 (Trakhtenbrot et al. 2015); 
and AGN correlation lengths of 7$h^{-1}$Mpc ($\sim$10\amin) at z$\sim$1-2, Allevato et al. 2011).
However, C-COSMOS only covered $\frac{1}{4}$ of COSMOS at $\sim$160 ks depth, 
plus 0.5 deg$^2$ at $\sim$80 ks depth (Fig. \ref{tiling}, green squares). 

We present here the \chandra\ \leg\ survey\footnote{Throughout the paper we use the term C-COSMOS to refer to the original survey of the inner field, and the name \chandra\ \leg\ survey to refer to the full, combined survey, including the new data presented here.}, 
the combination of the old C-COSMOS survey with 2.8 Ms of new \chandra\ ACIS-I (Garmire et al. 2003) observations (56$\times$50 ks pointings) approved 
during \chandra\ Cycle 14 as an X-ray Visionary Project (PI: F. Civano; program ID 901037).  
\leg\ uniformly covers the $\sim$1.7 deg$^2$ COSMOS/{\em HST} field at $\sim$160~ksec depth, expanding on the 
deep C-COSMOS area (dashed green square in Fig. \ref{tiling}) by a factor of $\sim$3 at $\sim$3$\times$10$^{-16}$ \cgs (1.45 vs 0.44 deg$^2$), for a total area covered of $\sim$2.2 deg$^2$.

This paper is the first in a series and presents the main properties of the survey and the X-ray point source catalog to be followed by a paper on the multiwavelength identification of the X-ray sources by Marchesi et al. (in preparation). 
In Section 2, we present the observations and tiling strategy. In Section 3, we detail all the steps of the data processing, including astrometric corrections, exposure and background map production. The data analysis procedure is instead described in Section 4, with some references and comparison with the one adopted for C-COSMOS as explained in P09. The point source catalog and the source properties are presented in Section 4.1. Last, in Section 5 and 6, the survey sensitivity and the number counts in both soft and hard band, and also dividing the sources in
obscured and unobscured, are presented.

We assume a cosmology with H$_0$ = 71~km~s$^{-1}$~Mpc$^{-1}$, $\Omega_M$ = 0.3 and $\Omega_{\Lambda}$= 0.7, and magnitudes are reported in the AB system if not otherwise stated. 
Throughout this paper, we make use of {\it J}2000.0 coordinates.  The data analysis is performed in three X-ray bandpasses 
0.5-2 keV (soft band, S), 2-7 keV (hard band, H), and 0.5-7 keV (full band, F), while sensitivity and fluxes have been computed in the 0.5-2, 2-10 and 0.5-10 keV bands for an easy comparison with other works in the literature.

\section{Observations}

\begin{figure}[t]
\centering
\fbox{\includegraphics[width=0.5\textwidth]{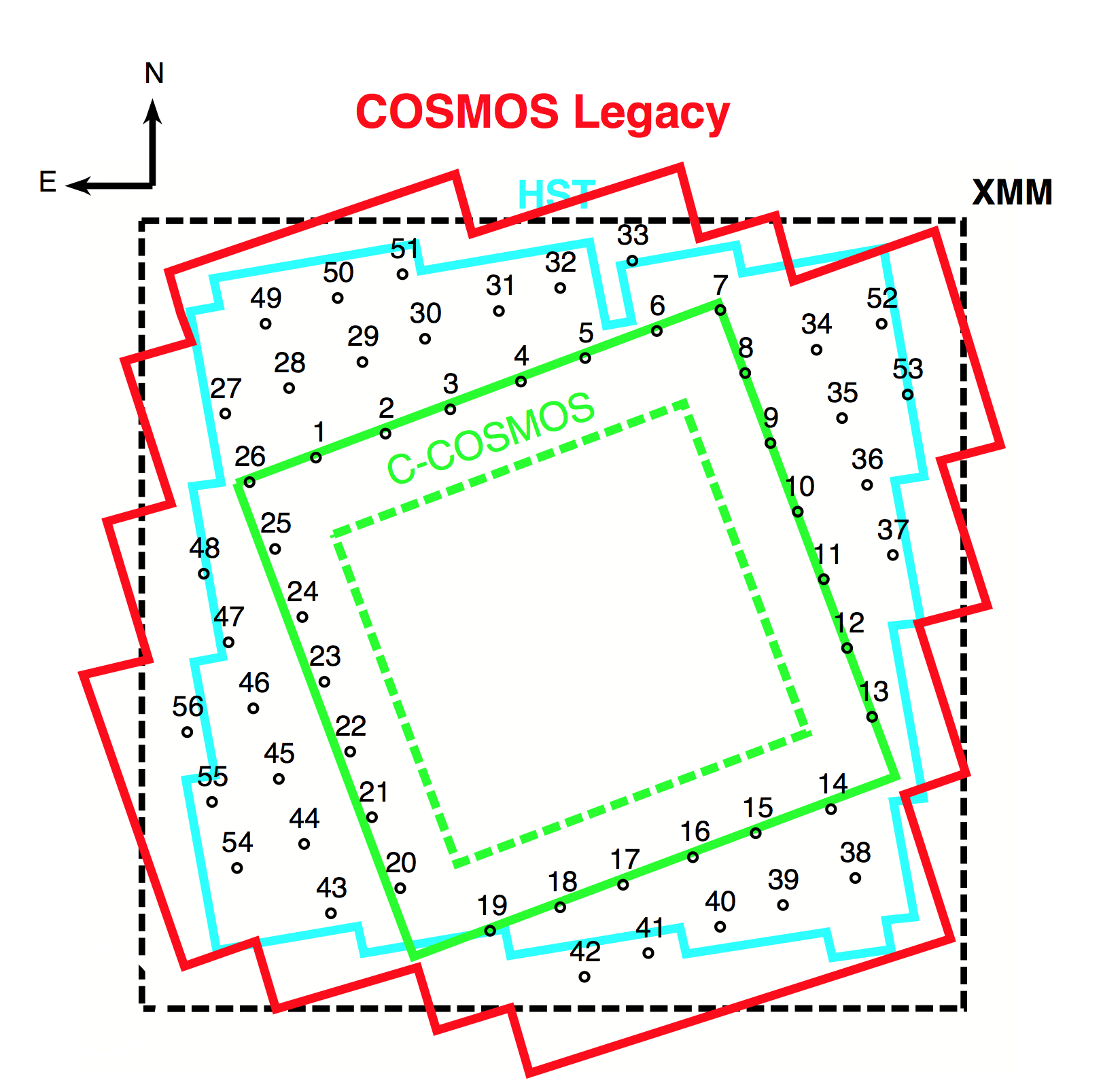}}
\caption{\leg\ tiling (red) compared to the area covered by HST (cyan), C-COSMOS (green solid: total area; green dashed: deeper area) and XMM-COSMOS (black). The ordering numbers of new observations are marked (See Table \ref{observation_list}).}
\label{tiling}
\end{figure}

The half-a-field shift tiling strategy was designed in order to cover uniformly, in depth and point spread function size, the COSMOS {\it Hubble} area (cyan outline in Fig. \ref{tiling}; Scoville et al. 2007b), by combining the old C-COSMOS observations (green outline in Fig. \ref{tiling}) with the new \chandra\ ones (red outline in Fig. \ref{tiling}). To achieve this, 56 ACIS-I pointings  (numbered black points in Fig. \ref{tiling}) were used, 11 of which were scheduled as two or more separate observations because of satellite constraints, for a total of 68 pointings. Moreover, the observing roll angle was constrained to be within 70$\pm$20 degree or 250$\pm$20. 
We summarize the main properties of the new \chandra\ COSMOS Legacy observations in Table \ref{observation_list}. 

The observations took place in four blocks: November, 2012 to January, 2013;  March to July, 2013; October, 2013 to January, 2014; and March, 2014. 
The mean net effective exposure time per field was 48.8 ks, after all the cleaning and reduction operations (see Section \ref{sec:process}). The maximum exposure was 53 ks (observation 15227) while the minimum exposure was 45.2 ks (combined observations 15208 and 15998).

The sequence of the observations was designed to start from the N-E top corner tile of C-COSMOS moving towards W and proceeding clockwise around the central C-COSMOS area, in such a way that the outer frame of the C-COSMOS survey overlaps with the inner frame of the new \chandra\ observations. The tiling number and the total area covered is shown in Figure \ref{tiling}. 

\begin{figure}[t]
\centering
\includegraphics[width=0.5\textwidth]{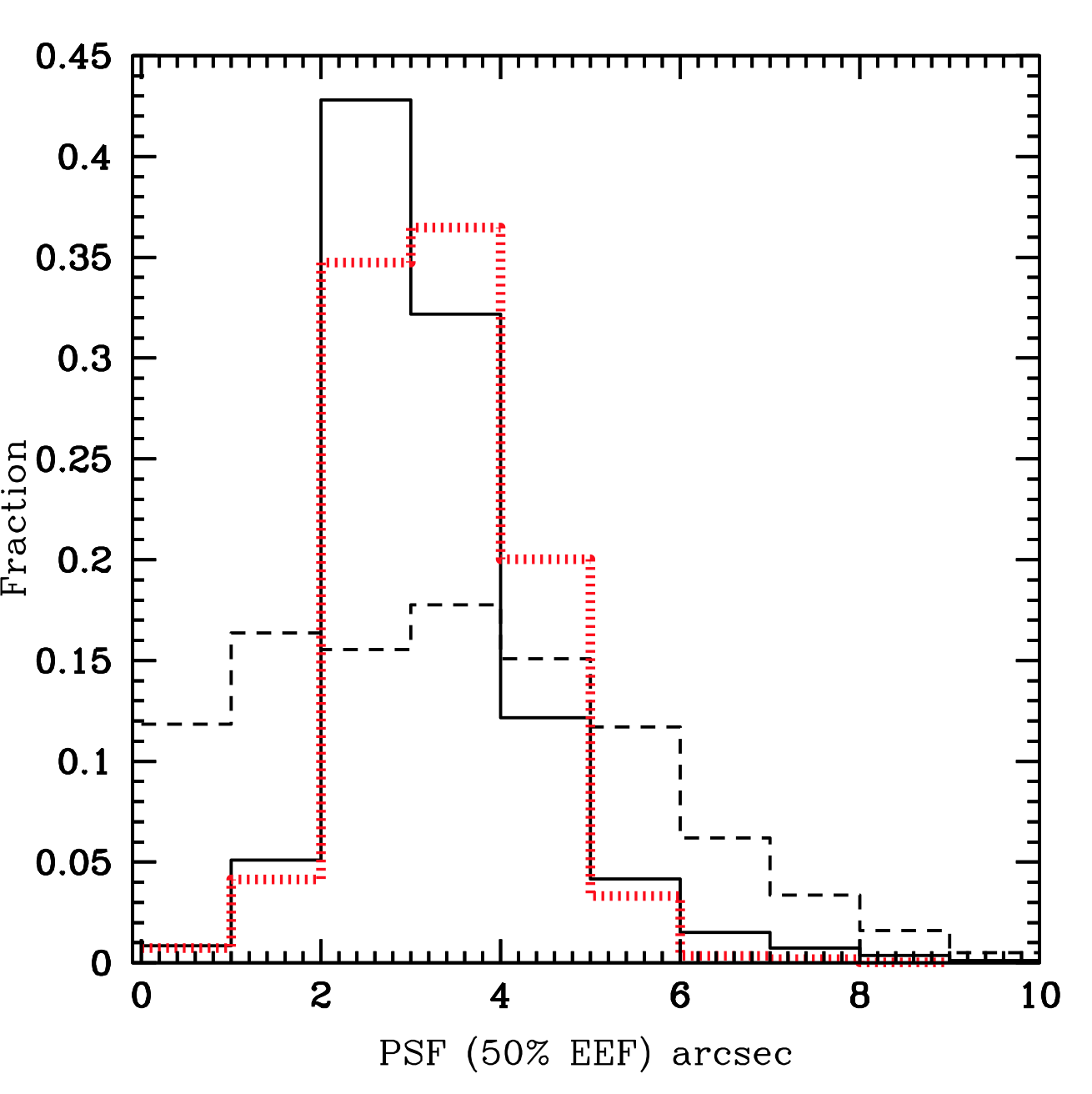}
\caption{Normalized distribution of the combined point spread function (50\% of EEF in 0.5-7 keV) size in arcsecond measured in \leg\ (solid histogram) and in a single pointing survey (dashed line). In red, the distribution of the combined PSF (the mean value) for all the detected sources.}
\label{psf_size}
\end{figure}

Using this tiling strategy we achieve an approximately uniform combined point spread function (PSF) across the survey. The mean combined PSF width (size at 50\% of the encircled energy fraction, EEF, in the 0.5-7 keV band; see Section \ref{sec:sensitivity} for details on the PSF maps), weighted on the exposure, peaks at around 3\arcsec\ (see Figure \ref{psf_size}). As shown in Figure \ref{psf_size}, 80\% of the field has a PSF in the range 2\arcsec-4\arcsec. As a comparison, in a single-pointed survey (regardless of exposure time), the PSF size distribution is flat, and although $\sim$30\% of the field has a PSF $<$2\arcsec\, the PSF can reach a substantially larger size ($>$ 4\arcsec) in 40\% of the field.

\begin{table*}[t]
\scriptsize
\centering
\caption{COSMOS Legacy Survey (CLS) Observation Summary}
\begin{tabular}{ccccccc}
\hline
\hline
Field$^a$ & Obs. ID & RA & Dec.& Date & Exp. time & Roll\\
& & & & & second & (deg)\\
\hline
CLS\_1 &	15207 & 150.544451 &  2.499045 &  2012-11-25  &  14883 & 70.2\\   
&	15590 & 150.544402 &  2.499094 &  2012-11-23  &  14893 & 70.2\\   
&	15591 & 150.544454 &  2.499065 &  2012-11-25  &  19828 & 70.2\\   
CLS\_2 &	15208 & 150.415643 &  2.543225 &  2012-12-07  &  22985 & 70.2\\   
&	15598 & 150.415625 &  2.543213 &  2012-12-08  &  22193 & 70.2\\   
CLS\_3 &	15209 & 150.295749 &  2.588083 &  2012-12-03  &  23775 & 70.2\\   
&	15600 & 150.295747 &  2.588106 &  2012-12-05  &  21795 & 70.2\\   
CLS\_4 &	15604 & 150.164741 &  2.639752 &  2012-12-10  &  20988 & 70.2\\   
&	15210 & 150.164738 &  2.639709 &  2012-12-16  &  24365 & 70.2\\	  
CLS\_5 &	15211 & 150.045569 &  2.682903 &  2012-12-13  &  23572 & 70.2\\   
&	15605 & 150.045586 &  2.682879 &  2012-12-15  &  21801 & 70.2\\	  
CLS\_6 &	15212 & 149.913425 &  2.732850 &  2012-12-21  &  25249 & 70.2\\   
&	15606 & 149.913418 &  2.732845 &  2012-12-23  &  25219 & 70.2\\   
CLS\_7 &	15213 & 149.796052 &  2.772968 &  2013-01-01  &  49435 & 62.2\\   
CLS\_8 &	15214 & 149.751287 &  2.655331 &  2013-01-03  &  45983 & 61.75\\  
CLS\_9 &	15215 & 149.704144 &  2.525446 &  2013-01-07  &  49437 & 63.2\\   
CLS\_10 & 15216 &149.654208 &  2.399733 &  2013-01-16  &  46459 & 56.7\\   
CLS\_11 & 15217 &149.627509 &  2.272922 &  2013-03-23  &  46057 & 265.2\\  
CLS\_12 & 15218 &149.584767 &  2.145874 &  2013-03-22  &  46475 & 265.2\\  
CLS\_13 & 15219 &149.538688 &  2.017596 &  2013-03-30  &  49432 & 261.6\\  
CLS\_14 & 15220 &149.614659 &  1.846399 &  2013-04-04  &  49924 & 60.1\\   
CLS\_15 & 15221 &149.753949 &  1.801935 &  2013-04-10  &  49431 & 58.2\\   
CLS\_16 & 15222 &149.870306 &  1.757718 &  2013-04-04  &  49407 & 60.0\\   
CLS\_17 & 15223 &149.999623 &  1.706079 &  2013-04-17  &  50905 & 55.2\\   
CLS\_18 & 15224 &150.115609 &  1.664373 &  2013-04-19  &  49426 & 55.2\\   
CLS\_19 & 15225 &150.245495 &  1.621716 &  2013-04-05  &  49631 & 59.8\\   
CLS\_20 & 15226 &150.411336 &  1.697830 &  2013-06-21  &  49428 & 250.2\\  
CLS\_21 & 15227 &150.463753 &  1.829216 &  2013-05-02  &  53051 & 50.2\\   
CLS\_22 & 15228 &150.504029 &  1.950647 &  2013-04-30  &  49432 & 50.2\\   
CLS\_23 & 15229 &150.551660 &  2.080265 &  2013-05-10  &  49012 & 52.2\\   
CLS\_24 & 15230 &150.592692 &  2.199969 &  2013-05-08  &  49429 & 52.2\\   
CLS\_25 & 15231 &150.642972 &  2.325853 &  2013-05-13  &  48446 & 51.0\\   
CLS\_26 & 15232 &150.690403 &  2.449284 &  2013-05-16  &  35085 & 50.6\\   
&	 15649 &150.690409 &  2.449315 &  2013-06-03  &  15251 & 50.65\\  
CLS\_27 & 15233 &150.734924 &  2.575875 &  2013-05-21  &  46476 & 50.20\\  
CLS\_28 & 15234 &150.616710&   2.623373&   2013-05-22 &   5439 & 50.20\\   
       & 15653  &150.594589 &  2.629150 &  2014-01-16  &  44895 &  58.21\\  
CLS\_29 & 15235 &150.480833 &  2.671101 &  2013-06-01  &  49440 & 48.31\\  
CLS\_30 & 15236 &150.364822 &  2.714260 &  2013-06-01  &  49435 & 48.23\\  
CLS\_31 & 15237 &150.228563 &  2.765929 &  2013-06-08  &  25246 & 50.65\\  
&	 15655 &150.228550 &  2.765907 &  2013-06-10  &  24466 & 50.65\\  
CLS\_32 & 15238 &150.114727 &  2.808579 &  2013-06-09  &  49429 & 50.65\\  
CLS\_33 & 15239 &149.981160 &  2.858739 &  2013-06-11  &  49430 & 50.20\\  
CLS\_34 & 15240 &149.617459 &  2.695912 &  2013-10-15  &  48450 & 77.09\\  
CLS\_35 & 15241 &149.593992 &  2.566795 &  2014-03-28  &  48600 & 260.21\\  
CLS\_36 & 15242 &149.547344 &  2.443553 &  2013-06-22  &  49432 & 50.20\\  
CLS\_37 & 15243 &149.499113 &  2.312060 &  2013-07-05  &  47985 & 50.20\\  
CLS\_38 & 15244 &149.547442 &  1.723759 &  2014-01-21  &  47461 & 53.21\\  
CLS\_39 & 15245 &149.680897 &  1.673516 &  2014-01-23  &  49437 & 53.21\\  
CLS\_40 & 15246 &149.796705 &  1.629086 &  2013-10-22  &  48850 & 75.21\\  
CLS\_41 & 15247 &149.953115 &  1.578784 &  2014-03-18  &  49545 & 267.21\\  
CLS\_42 & 15248 &150.047419 &  1.537353 &  2013-11-13  &  49438 & 71.61\\  
CLS\_43 & 15249 &150.516513 &  1.655025 &  2013-11-29  &  45635 & 70.21\\  
CLS\_44 & 15250 &150.566083 &  1.783479 &  2013-12-12  &  49315 & 70.21\\  
CLS\_45 & 15251 &150.612991 &  1.904134 &  2013-12-03  &  29702 & 67.91\\  
       & 16544 &150.613008 &  1.904121 &  2013-12-04  &  19830 & 67.91\\  
CLS\_46 & 15252 &150.660018 &  2.034094 &  2013-12-14  &  49434 & 70.21\\  
CLS\_47 & 15253 &150.707972 &  2.162869 &  2014-01-28  &  49132 & 53.21\\  
CLS\_48 & 15254 &150.753963 &  2.289683 &  2014-01-29  &  49139 & 53.21\\  
CLS\_49 & 15255 &150.661405 &  2.741395 &  2014-03-24  &  49435 & 260.21\\  
CLS\_50 & 15256 &150.504801 &  2.795740 &  2014-01-13  &  49943 & 59.21\\  
CLS\_51 & 15257 &150.384246 &  2.838987 &  2014-01-04  &  49435 & 61.85\\  
CLS\_52 & 15258 &149.497504 &  2.746858 &  2014-01-01  &  49432 & 62.27\\  
CLS\_53 & 15259 &149.451159 &  2.620733 &  2014-01-27  &  49435 & 53.21\\  
CLS\_54 & 15260 &150.690908 &  1.740589 &  2014-01-05  &  22793 & 60.21\\  
       & 16562 &150.690921 &  1.740576 &  2014-01-25  &  26736 & 60.21\\  
CLS\_55 & 15261 &150.736977 &  1.863191 &  2014-01-18  &  46474 & 59.21\\  
CLS\_56 & 15262 &150.782957 &  1.992193 &  2014-01-12  &  50236 & 59.21\\  
\hline
\end{tabular}
\label{observation_list}
\end{table*}

\begin{figure}[t]
\centering
\includegraphics[width=0.5\textwidth]{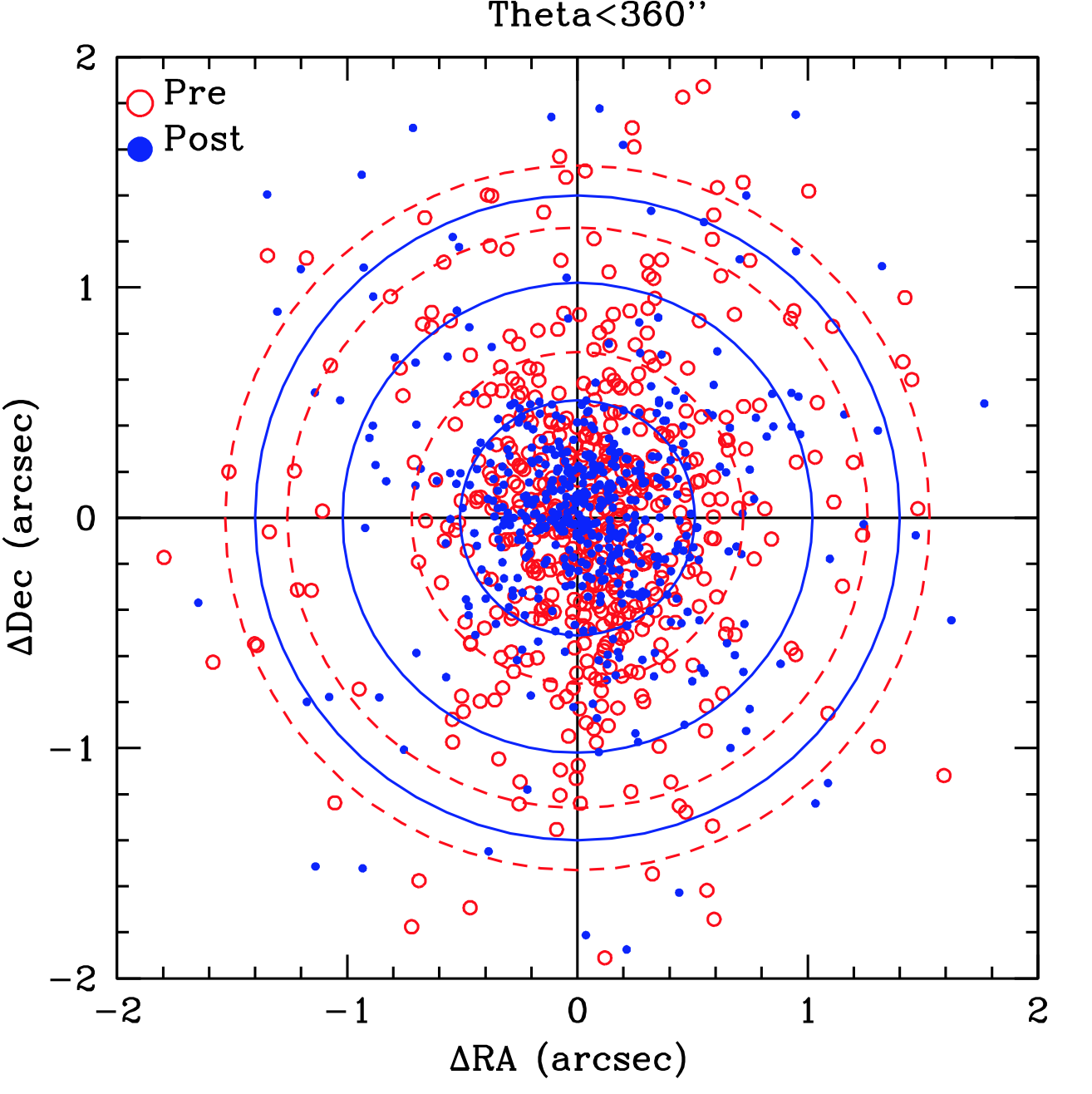}
\caption{The X-ray to I-band separation ($\Delta$RA, $\Delta$dec) in arcsecond for 
X-ray sources within 6$^{\prime}$ from the aim point detected in each single observations before (red open circles) and after (blue solid circles) the aspect correction.
The circles encompass 68\%, 90\% and 95\% of the sources before (red dashed) and after (blue solid) the correction. }
\label{astrometry}
\end{figure}

\section{Data processing}
\label{sec:process}

The data reduction was performed following the procedures described in E09 for C-COSMOS, using standard \chandra\ CIAO 4.5 tools (Fruscione et al. 2006) and CALDB 4.5.9. We also reprocessed the 49 C-COSMOS observations in order to use them in concert with the new observations for source detection in the area where the new observations overlap with the old ones and to compute the sensitivity of the whole survey (see the comparison between fluxes in Section \ref{sec:source_fluxes}). 

We used the \texttt{chandra\_repro} reprocessing script, which automates the CIAO recommended data processing steps and creates new level 2 event files, applying the {\it VFAINT} mode for ACIS background cleaning to all the observations. 
We then performed the following steps before starting data analysis: astrometric correction and reprocessing of all the observations to a standard frame of reference using the new aspect solution (Section \ref{sec:astro}); mosaic and exposure map creation in three standard \chandra\ bands (Section \ref{sec:expo}): 0.5-7 keV, 0.5-2 keV and 2-7 keV; background map creation, using a two-components model to take into account both the cosmic background contribution and the instrumental one (Section \ref{sec:bkg}).

\subsection{Astrometric corrections}\label{sec:astro}
Even though \chandra\ data astrometry is accurate to 0.6\arcsec\ (at 90\% confidence, see Proposer User Guide\footnote{http://cxc.cfa.harvard.edu/proposer/POG/html/chap5.html\#tth\_fIg5.5} Chapter 5), in order to produce a sharp X-ray mosaic and to match the positions of X-ray sources with the optical catalog for which the positional accuracy is $\sim$0.2\arcsec\ (Capak et al. 2007, Ilbert et al. 2009, Laigle et al. submitted), we performed source detection on each individual observation to register them to a common optical astrometric frame. This work has been done on the new observations and also on the C-COSMOS outer frame fields overlapping with the new data. 
We generated a list of detected sources using the CIAO wavelet source detection tool \texttt{WAVDETECT} on each single observation binned at 1\arcsec\ and adopted a false-positive detection probability threshold corresponding to $\sim$10 spurious sources per field. Of the detected sources (on average 150 sources per field), we considered in each field those with significance $>$3.5$\sigma$ and within 360\arcsec\ from the aim point. In \chandra\ data, the positional accuracy of significant sources is $<$1\arcsec\ even at 10$^{\prime}$ off-axis and it is energy independent (K. Glotfelty private communication). Therefore, choosing sources within 6$^{\prime}$ of the aim-point provides a sample of sources with very good centroid estimate ($<$0.3\arcsec) for astrometric purposes. Using the CIAO tool \texttt{reproject\_aspect}, these sources were then compared to the CFHT {\it MegaCam} catalog of i-band selected sources (McCracken et al. 2012) with optical magnitudes in the range 18-23 AB {\it mag}. At least 4 sources in each field, not on the same side of the aim-point, are needed to compute meaningful rotational and translation transformations. In our analysis, we used on average 12 sources (up to 22 sources per field), with 75\% of the fields having more than 10 sources used to perform the reprojection.

With the corrected aspect solution, we reprocessed the level 1 data using \texttt{chandra\_repro} and performed the \texttt{WAVDETECT} detection again to compute the new separation between X-ray and optical positions. The resulting standard deviation on the shift computed from the detected sources within 6$^{\prime}$ is 0.36$^{\prime\prime}$ and 0.51$^{\prime\prime}$ on the RA and Dec, respectively. After matching all the X-ray fields to the same astrometric optical frame, 95\% of the X-ray sources used for the astrometry correction have a distance to their optical counterpart smaller than 1.4$^{\prime\prime}$, 10\% lower than the value before the correction (1.53$^{\prime\prime}$). The improvement in the position increases to 20\% when considering 90\% of the sources (1.26\arcsec\ to 1.02\arcsec) and 30\% when considering a smaller sample of 68\% of the sources (from 0.72\arcsec\ to 0.51\arcsec; see Figure \ref{astrometry}). This is consistent with, and slightly better than, what was found for C-COSMOS (see E09, Figure 6). 

\subsection{Exposure maps and data mosaic creation}
\label{sec:expo}

We created exposure maps in three bands using the standard CIAO procedure.
The spectral model used for the map creation is a single power-law with slope $\Gamma$=1.4\footnote{The choice of such a spectral slope is not only because of consistency with E09 and P09, but it is also the slope of the cosmic X-ray background (e.g., Hickox \& Markevitch 2006) and therefore well represents a mixed distribution of obscured and unobscured sources at the fluxes covered by \leg.} and Galactic absorption N${_H}$=2.6$\times$10$^{20}$ cm$^{-2}$ (Kalberla et al. 2005). Instrument maps, generated with \texttt{MKINSTMAP} for each CCD in each observation, were used as input files for the \texttt{MKEXPMAP} tool, which computes an exposure map for each CCD separately. These exposure maps were combined in a single exposure map for each observation using \texttt{DMREGRID} with a binning of 2 pixels. 

\begin{figure*}
  \centering
\fbox{  \includegraphics[width=\textwidth]{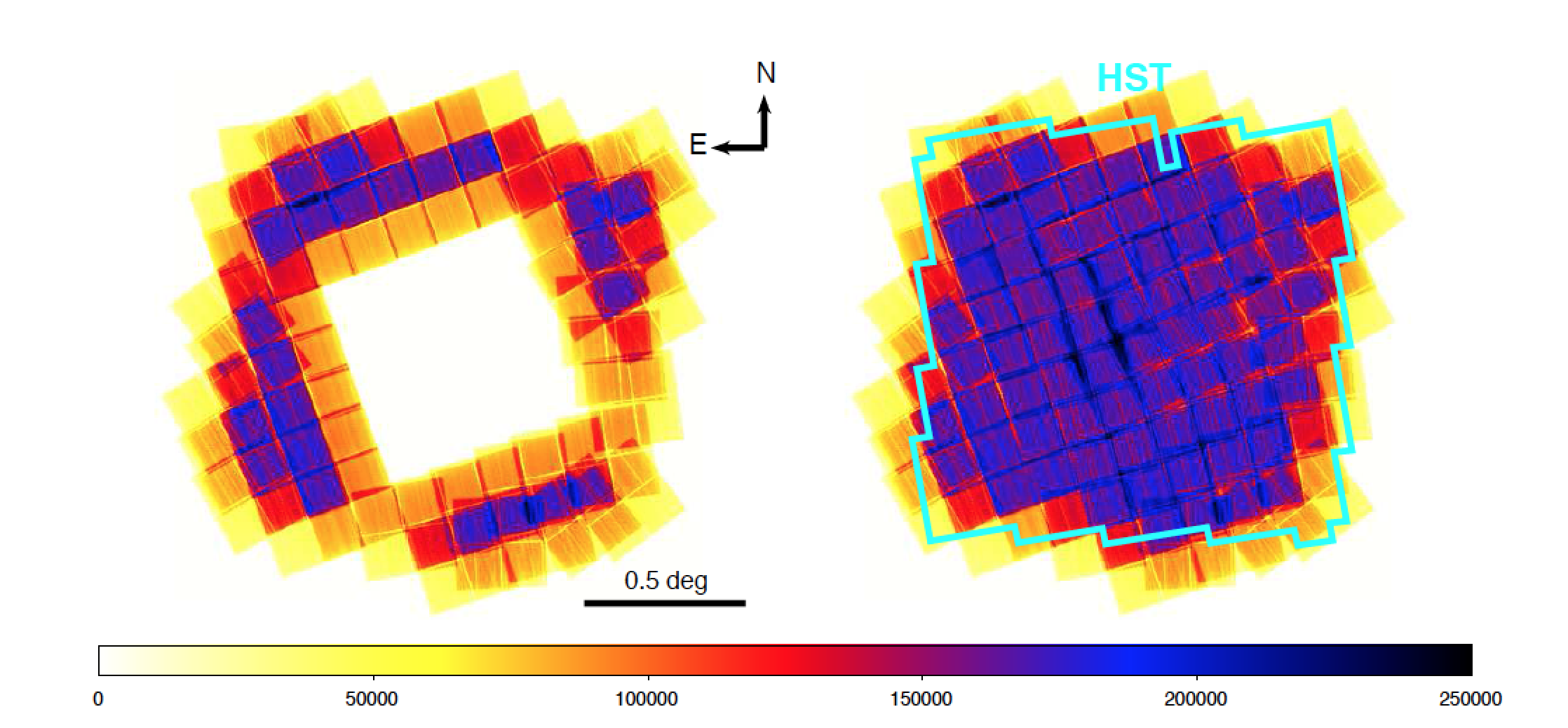}}
\caption{The mosaic of exposure maps for the new observations (left) and for the whole \leg\ survey (right) in the full band. The color bar gives the achieved effective exposure in units of seconds. We reached a uniform coverage of $\sim$160 ks over the full HST area (cyan polygon).}
\label{fig:exp}
\end{figure*}

Figure \ref{fig:exp} shows a composite image of the effective exposure time (in seconds) in the full band for both the new observations (left) and the whole \leg\ (right). As can be seen, the central 1.5 deg$^2$, covering almost entirely the HST area, have a uniform depth of $\simeq$160 ks.

\begin{figure*}
  \centering
\fbox.png{\includegraphics[width=\textwidth]{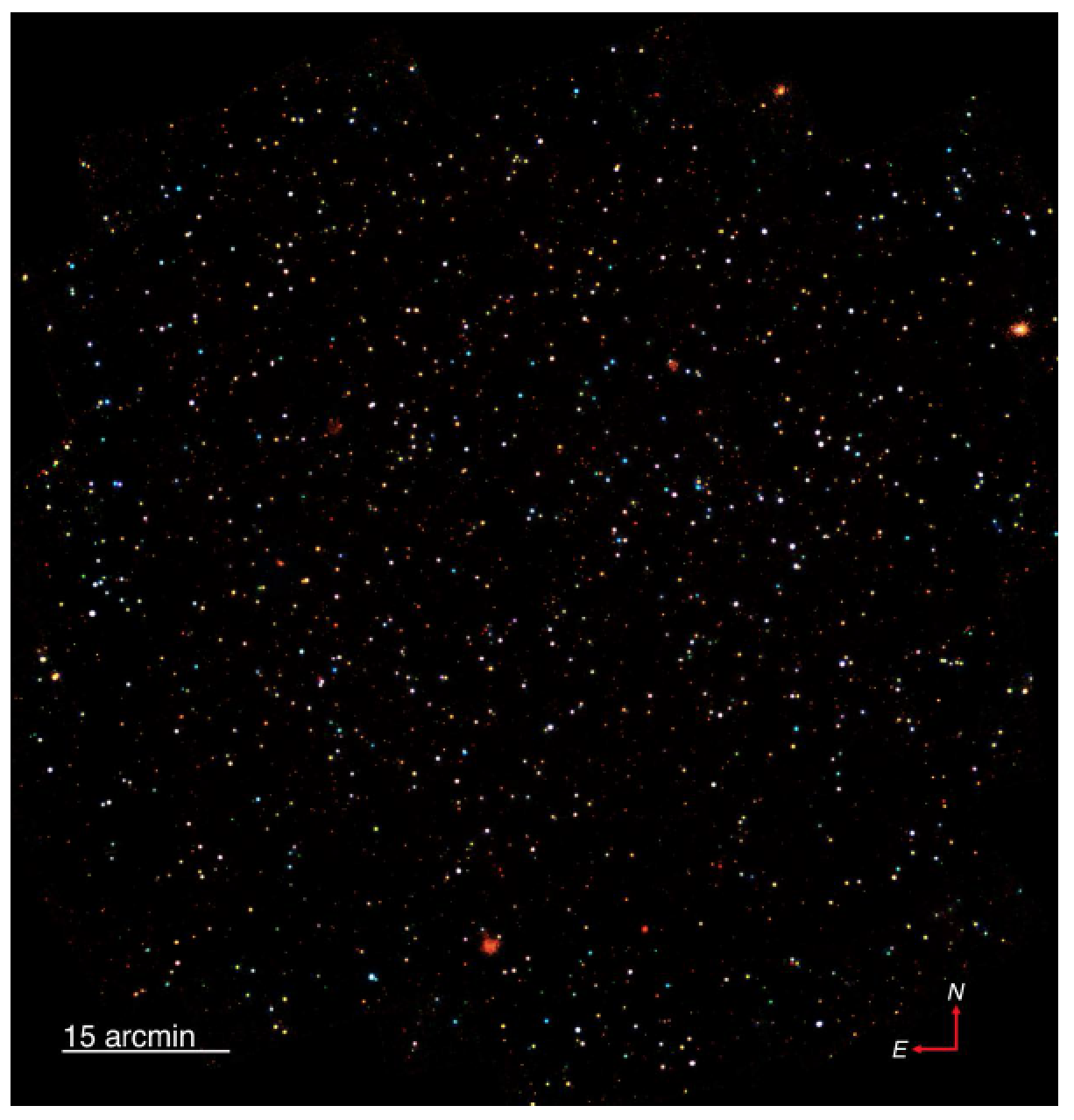}}
\caption{Three color image of the whole \leg\ field (0.5-2.0 keV, 2.0-4.5 keV, and 4.5-7.0 keV as red green and blue, respectively).}
\label{fig:mosaic}
\end{figure*}

The data mosaic image was created in three bands using the HEASoft \texttt{addimages} tool, which adds together a set of images using sky coordinates. In Figure \ref{fig:mosaic}, the three color image, created by combining the exposure corrected images in three non-overlapping bands (0.5-2.0 keV, 2.0-4.5 keV, and 4.5-7.0 keV as red green and blue, respectively) is shown. The combined image was then Gaussian smoothed with a 3 pixel radius. A filter was then applied to isolate sources from the background level, as well as to increase the contrast and color vibrancy of those sources. This process was repeated 3 times. 

\subsection{Background maps creation}
\label{sec:bkg}

The \chandra\ background consists of two different components: the cosmic X-ray background and a quiescent instrumental background due to interactions between the ACIS-I CCD detectors and high-energy particles. We followed the procedure described in Cappelluti et al. (2013) to create background maps, which we used for the selection of reliable sources in our detection procedure and for the computation of the sensitivity curves.

The background maps were computed for each observation separately in the full, soft and hard bands. We ran \texttt{WAVDETECT} with a threshold parameter \textit{sigthresh}=10$^{-5}$,  corresponding to $\sim$100 spurious sources per field (see Section \ref{sec:astro}), large enough to select also sources with significant signal only in stacked emission. We then removed these sources from the science images by excising a region corresponding to the source size (using a 3$\sigma$ value) as computed by the detection tool. We then uniformly distributed the remaining counts, rescaled by the ratio between the whole area of the observation and the area without the removed sources. These files were then used as initial background.

We then downloaded ``stowed background'' data from the \chandra\ archive\footnote{http://cxc.harvard.edu/ciao/threads/acisbackground/}. Stowed background files are particle-only background files and are obtained when the ACIS detector is out of the focal plane. These files were then rescaled using the procedure described in Hickox \& Markevitch (2006): we measured the ratio between the number of counts in our initial background ($C_{data}$) and in the stowed image ($C_{stow}$) in the energy range 9.5-12 keV. In this band, the effective area of \chandra\ is $\simeq$0 and consequently all the counts have a non-astrophysical origin. 

The stowed background, rescaled to our data by $C_{data}$/$C_{stow}$, was then subtracted from the initial background to obtain a first version of the cosmic X-ray background. The counts of this map were then renormalized using the exposure maps to create an exposure-corrected cosmic background. 

Finally, we performed a Monte Carlo simulation using the exposure-corrected cosmic X-ray background and the stowed background as input files. We simulated 1000 images for each of the two backgrounds using the IDL routine {\it poidev} to obtain a Poissonian realization of each map, and then we obtained our final homogeneous background map adding together the two mean simulated images. In order to use these maps for sensitivity computations and in our detection algorithm, a Gaussian smoothing (with a scale of 20 pixels) was applied to this final background map using the FTOOL \texttt{fgauss}. 

\begin{figure}[t]
  \centering
  \includegraphics[width=0.5\textwidth]{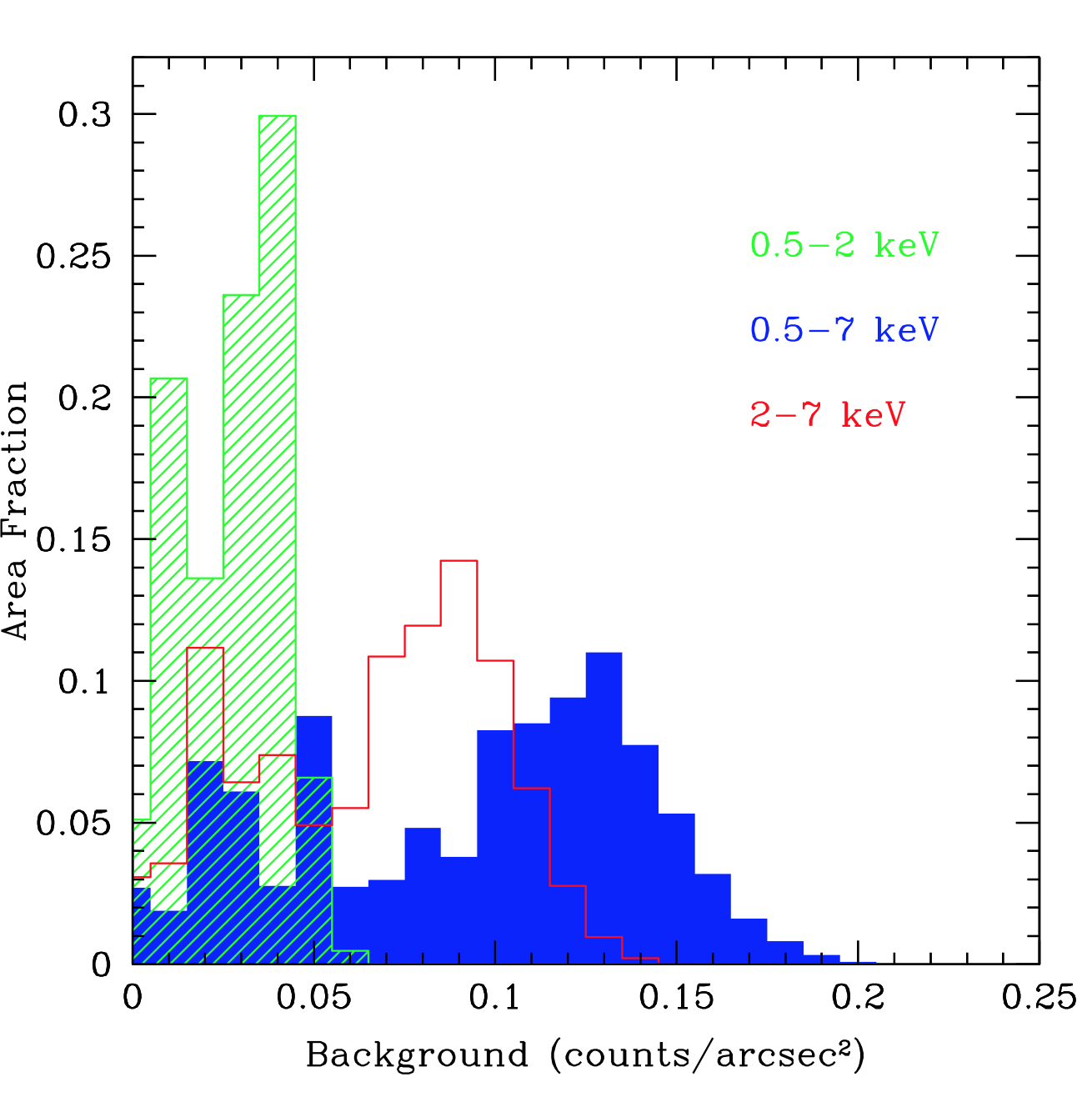}
\caption{Distributions of background counts per square arcsecond in the full (solid blue histogram), soft (shaded green histogram) and hard (empty red histogram) bands.}
\label{background_histo}
\end{figure}

The distribution of the computed background (in counts/arcsec$^2$) in the three bands is reported in Figure \ref{background_histo}. The overall background count distribution is consistent with the one found in C-COSMOS (see Figure 4 of P09). In the full band the main peak is at around 0.13 counts/arcsec$^2$ and this corresponds to the deepest part of the exposure. In C-COSMOS, the deep and shallow areas were roughly the same size and therefore the background distribution had two clear peaks of approximately the same height, while in \leg, the area with higher exposure is 3 times larger than the shallow area. This is represented in the background distribution as well. The number of background counts is consistent with the expectation for \chandra\ given the distribution of our exposure times.  

\section{Data Analysis: source detection and photometry}
\label{sec:process}

The analysis presented in the following focuses only on point sources. A parallel effort on the detection of extended sources will be presented by Finoguenov et al. (in preparation). To avoid contamination by extended sources, we used the XMM-COSMOS catalog of extended sources (Finoguenov et al. 2007, Kettula et al. 2013) and visually inspected all the brightest (L$_X >$ 10$^{41}$ \lum\ in 0.5-2 keV) ones to check if a point source is detected inside them by \chandra. 

Puccetti et al. (2009) extensively discussed and compared different source detection techniques concluding that the best procedure for C-COSMOS was a combination of \texttt{PWDetect} (Damiani et al. 1997) and the \texttt{Chandra Emldetect} (\texttt{CMLDetect}) Maximum Likelihood algorithm. 
As shown by P09 using extensive simulations, one of the strongest features of \texttt{PWDetect} is its ability to locate X-ray sources with extreme accuracy (0.02\arcsec$\pm$0.15\arcsec, P09 Table 1), while \texttt{CMLDetect} is the best tool to perform source photometry and derived source significance. The \leg\ survey shares the same tiling layout, exposure time per field and roll angle range of C-COSMOS, hence, we can follow the P09 procedure and use the same significance threshold for source detection. 

The original version of \texttt{CMLDetect}, called \texttt{emldetect} (Cruddace et al. 1988, Hasinger et al. 1993), is part of the \xmm\ SAS package and is based on a code originally developed for \textit{ROSAT} data. \texttt{CMLDetect} has been adapted to run on \chandra\ data by replacing the \xmm\ PSF library with the \chandra\ one (see Krumpe et al. 2015 for another application of \texttt{CMLDetect}). Moreover, this new tool can also work with different PSFs simultaneously. 

 \texttt{PWDetect} was developed to properly treat \chandra\ data with PSF varying across the field and it is based on the wavelet transform (WT) of the X-ray image. A WT is the convolution of an image with a ``generating wavelet'' kernel which depends on position and length scale (a free parameter). For this survey, and for \chandra\ data in general, the length scale varies from 0.5\arcsec\ to 16\arcsec\ in steps of $\sqrt{2}$. These steps cover all possible \chandra\ PSFs (the largest are those at large off axis angle $\theta_{i}$). Both radial and azimuthal PSF variations are accounted for by \texttt{PWDetect}, which first assumes a Gaussian PSF and then corrects by a PSF shape factor, calibrated with respect to source positions on the CCD.

\texttt{PWDetect} works on stacked observations only if co-aligned (same aim point and roll-angle), as is the case for 11 of our fields which are observations split into multiple parts. Therefore, \texttt{PWDetect} was run on each of our new 56 fields setting the detection limit to 3.8$\sigma$ corresponding to a probability of a spurious detection to $\simeq$10$^{-4}$ with the aim of creating a large catalog of detections to be fed to \texttt{CMLDetect}. Also, given that the outer frame of C-COSMOS overlaps with the new survey, we run \texttt{PWDetect} on 20 old fields (fields 1-1 to 1-6, 1-6 to 6-6, 6-6 to 6-1 and last 6-1 to 1-1 as in Table 3 of E09). For overlapping regions between different pointings, we performed a positional cross correlation (using a 2\arcsec\ radius) and if a source was detected in more than one field, we chose the position of the source at the smallest $\theta_{i}$, i.e. the one with the best PSF. We performed a visual inspection of all the 
sources having multiple matches within 5\arcsec. About 90\% of the pairs in the range 2--5\arcsec\ were actually false detections, mainly caused by PSF tail detection of bright sources. 

The positions obtained with \texttt{PWDetect} were then fed as input to \texttt{CMLDetect}, to obtain photometric information and significance for each source. We ran \texttt{CMLDetect} allowing the detection only of point-like sources.
\texttt{PWDetect} can be used to obtain net counts, rates and fluxes, but we opted to use \texttt{CMLDetect} because it can work on a mosaic, while \texttt{PWDetect} cannot. Moreover, P09 has shown that \texttt{PWDetect} count rates are systematically less accurate than those of \texttt{CMLDetect} (the median ratio between the output detected and input simulated count rates ranges from 86 to 94\% for \texttt{PWDetect} versus 97 to 105\% for \texttt{CMLDetect}, independently from the energy). \texttt{CMLDetect} performs a simultaneous maximum likelihood PSF fitting for each input candidate source, previously obtained using \texttt{PWDetect}, to all images at each position and, working on a mosaic, can provide a refined position of the source and count rates. This procedure was run in three bands: full (0.5-7 keV), soft (0.5-2 keV) and hard (2-7 keV). With the goal of not missing close pairs, we run \texttt{CMLDetect} allowing to slightly change the 
input position provided by \texttt{PWDetect}.

The best-fit maximum likelihood parameter in \texttt{CMLDetect},  DET\_ML, is related to the Poisson probability that a source candidate is a random fluctuation of the background ($P_{random}$), as follows:

\begin{equation}
 DET\_ML = -ln(P_{random}).
\end{equation}

As a consequence, sources with small values of  DET\_ML have high values of $P_{random}$ and are then likely to be background fluctuations. We chose a threshold significance value of 2 $\times$ 10$^{-5}$, that corresponds to DET\_ML=10.8, i.e., a source needs to have DET\_ML$>$10.8 in at least one of the three bands to be included in the final catalog. This value is the same used in C-COSMOS and represents the best compromise between completeness and reliability as shown by P09 in Figure 11 and 12. 75\% of the sources detected by \texttt{PWDetect} in a single field with DET\_ML$>$10.8 and fed to \texttt{CMLDetect}, were found to be above the threshold in output. 

To improve the final completeness of the catalog, we also search for less significant sources, up to about 100 times higher $P$, which corresponds to a threshold  DET\_ML=6. Similarly to what was done in C-COSMOS, sources with DET\_ML in the range 6 to 10.8 and are only considered in this catalog if these have DET\_ML$>$10.8 in another band. Sources with DET\_ML$<$6 are considered undetected.

P09 performed extensive Monte Carlo simulations for C-COSMOS to both test the detection and photometry strategy as well as to determine the completeness and reliability of the source catalog at the chosen DET\_ML threshold. Given that \leg\ is the scaled up version of C-COSMOS (in area and exposure), the same analysis was followed, therefore we infer that the  completeness and reliability of the catalog are the same. Therefore, the chosen DET\_ML threshold implies a completeness of 87.5\% and 68\% for sources with at least 12 and 7 full band counts, of 98.2\% and 83\% for the soft band, 86\% and 67\% for the hard band.  At this significance level and the same count limits, the reliability is $\sim$99.7\% for the three bands.

\subsection{Point source catalog}
\label{sec:catalog}

\subsubsection{Source numbers}

We positionally matched the three single-band, \texttt{CMLDetect} output catalogs (including all the sources to  DET\_ML=6) to one another using a cross-correlation radius of 3\arcsec. 
We first matched the full band detected source catalog to the soft band one, then the full with the hard band catalog and finally the soft and hard band one. We performed a visual inspection of the whole sample, and we also made use of the catalog of optical/IR identifications (presented in a companion paper, Marchesi et al. in press), to solve ambiguous cases. After the visual inspection, we found that $<$1\% of the matches are actually fake associations and all related to sources at the outer edges of the survey with rather wide point spread function, therefore with large positional error.
Overall, the mean (median) separation between detections of the same source in two different bands is 0.43\arcsec\ (0.23\arcsec) for full to soft and 0.41\arcsec\ (0.23\arcsec) for full to hard, with 90\% of the matches within 1\arcsec. For soft to hard matches, the mean (median) separation is instead 0.73\arcsec\ (0.56\arcsec), with $\sim$80\% within 1\arcsec. The source position is determined in the full band for all the sources detected in the full band; if a source is not detected the full band, the soft band position is used. The hard band position is used for sources detected in the hard band only.

In Table \ref{tab:x_data}, we report the total number of new sources for each combination of bands, while in Table \ref{tab:x_single} we report the number of sources detected in each band at the two adopted thresholds (DET\_ML$>$10.8 and 6$<$DET\_ML$<$10.8). The number of detections with DET\_ML$>$10.8 in at least one of three X-ray bands is 2273. 
The number of expected spurious sources with DET\_ML$>$10.8 is reported in each band for two count limits in Table \ref{tab:spurious}.

In the area where the new data overlap with the outer C-COSMOS frame, the exposure time is now double the previous mean exposure time (142 ks versus 72 ks), and 385 new sources are detected in addition to the 694 sources already in E09. For the last 694 sources with doubled exposure time, 676 have been detected in the new data as well. The eighteen C-COSMOS sources  not detected in the new data had DET\_ML values in E09 in the three bands close to the threshold (DET\_ML $<$15); moreover, 10 of them were detected only in 2 out of 3 bands in E09 and the remaining eight were detected only in 1 band.  

In Table \ref{tab:x_data}, we include the number of sources in each combination of bands for the C-COSMOS area including the new data and also in parentheses the number of sources as in E09. The same old and new numbers are included in Table \ref{tab:x_single}. In this paper we provide also an updated catalog of the C-COSMOS sources with larger exposure in the total data. Among the 676 C-COSMOS sources with new data, only $\sim$1.5\%, $\sim$2\% and $\sim$3\% in the full, soft and hard band, respectively, have a DET\_ML value which is below the threshold in the combined data while it was above the 10.8 DET\_ML threshold in the C-COSMOS catalog, confirming the reliability of the detection method, and the consistency between the analysis performed in E09 and P09 and the one performed here. The actual fraction of sources with DET\_ML lower in the combined dataset than in C-COSMOS is 14\% in the full and 10\% in the soft and hard bands. On average, sources with lower DET\_ML in the new dataset are in an area of the field where the ratio $(exp\_new-exp\_old)/exp\_old$ is 40\% lower than the average ratio of the sources in the catalog, therefore the discrepancy could be explained with source variability.

The total number of sources summing the two datasets is reported in the last column of Table \ref{tab:x_data}.  Adding the new observations, we more than double the sample with respect to C-COSMOS, obtaining a catalog of 4016 sources, the largest sample of X-ray sources homogeneously detected and with uniform multiwavelength data (see Section \ref{summary} for a discussion and Marchesi et al. in press). In comparison, other contiguous surveys with similar area in the literature have about 20\% fewer sources than \leg\ (see 3362 sources in Stripe 82 by LaMassa et al. 2013a,b and 2015; 3293 in X-Bootes by Murray et al. 2005; 2976 in X-DEEP2 by Goulding et al. 2012). 

\begin{figure}[t]
\centering
\includegraphics[width=0.5\textwidth]{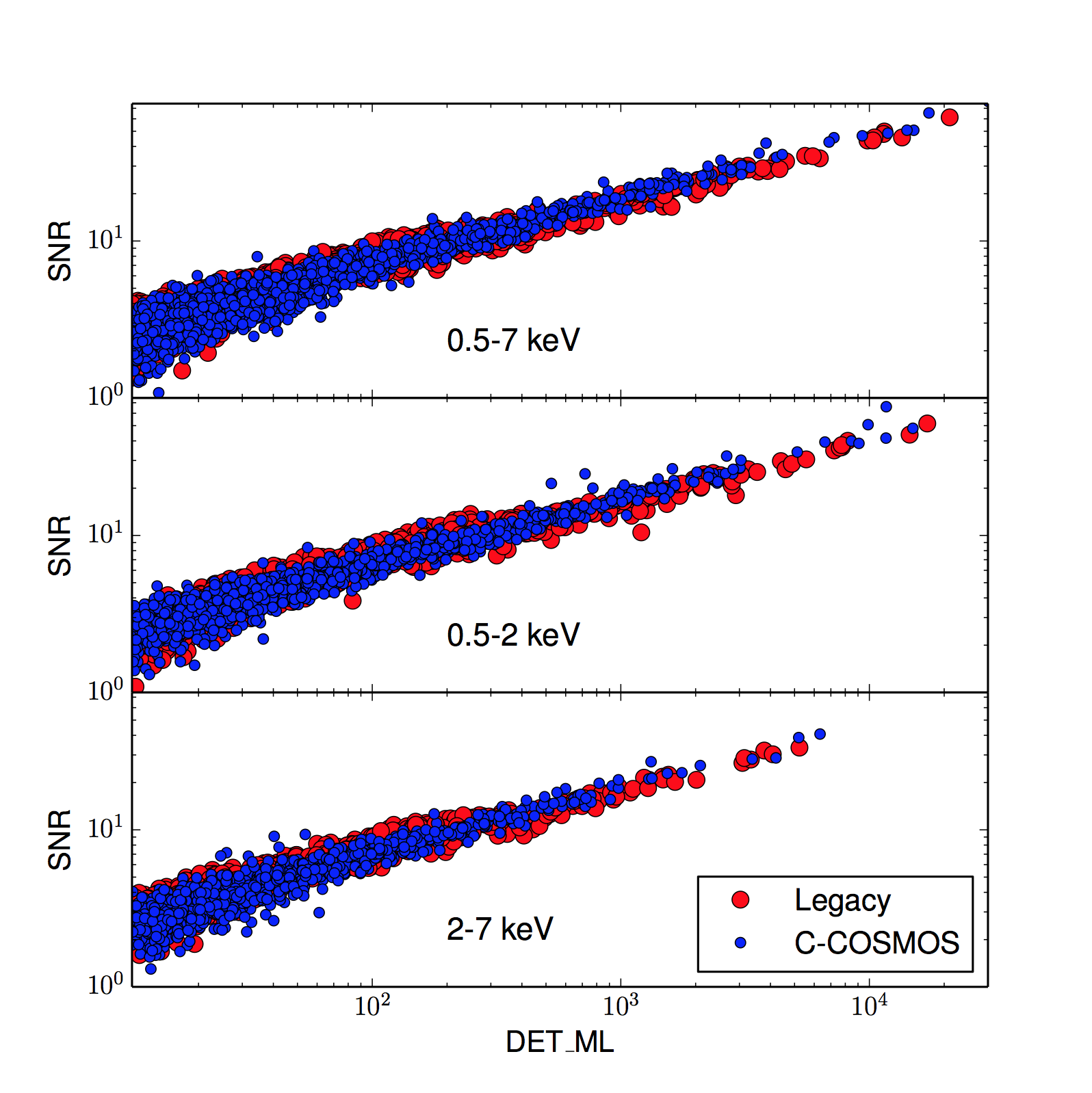}
\caption{Signal-to-noise ratio as a function of DETML for sources detected in three bands. The new \chandra\ sources are plotted as red circles, the C-COSMOS sources as blue ones. We plot only sources with DETML$>$10.8.}
\label{fig:det-snr}
\end{figure}

In Figure \ref{fig:det-snr}, we show the signal-to-noise ratio (SNR = count rate/count rate error) as a function of the DET\_ML for the new sources with  DET\_ML$>$10.8. In excellent agreement with the finding in C-COSMOS, the SNR increases smoothly with increasing  DET\_ML, with a dispersion of a factor of 2 at both low and high  DET\_ML values.

\subsubsection{Source positional errors}

\begin{figure}[t]
\centering
\includegraphics[width=0.5\textwidth]{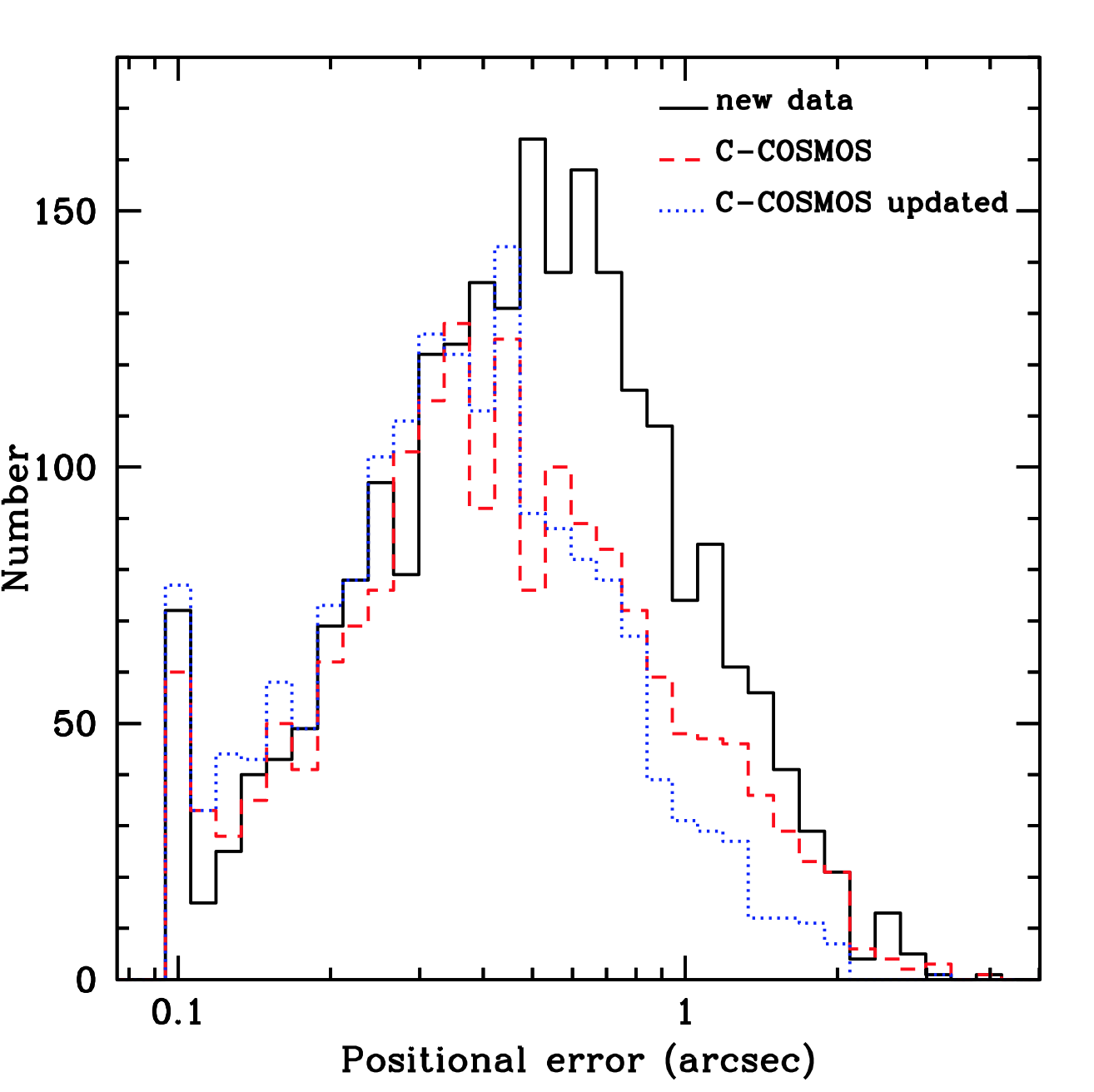}
\caption{Positional error distribution for the new \leg\ data (black solid line), the original C-COSMOS (red dashed line) and the updated C-COSMOS (blue dotted line).}
\label{pos_error}
\end{figure}

To compute the positional errors associated with the X-ray centroids given in the catalog ($\sqrt{\sigma^2_{R.A.}+\sigma^2_{dec}}$), we followed the prescription of P09 defining $err\_pos=r_{PSF}/\sqrt{S}$, where S is the number of net (i.e. background subtracted) source counts in the full band in a circular region of radius $r_{PSF}$ containing 50\% of the encircled energy in the observation where the source is at the smallest off-axis angle.  The positional errors are generally in very good agreement with those resulting from \textit{CMLDetect}. In Figure \ref{pos_error}, the positional error distribution is presented for all the new sources (black solid), the old C-COSMOS sources (red dashed) and the updated C-COSMOS distribution (blue dotted). 
The sources plotted in the lowest bin are those with positional error values actually smaller than 0.1\arcsec, which we set to 0.1\arcsec, consistently with the work done in P09. These sources with small positional error are just very bright objects (with $\sim$240 mean full band counts; see next section).

The peak of the new sources distribution is $\sim$0.6\arcsec\ and 85\% of the sources have a positional error $<$1\arcsec, while C-COSMOS source distributions peaks at around 0.4\arcsec. This difference (the somewhat larger positional errors for the sources detected with the new data than for those detected in C-COSMOS) is due to the fact that, as shown in Fig. \ref{counts_histo}, the net counts distribution for the sources in the new data peaks at a lower value than for the C-COSMOS sources (therefore giving a smaller denominator in the formula of the positional error).

\subsubsection{Source counts and fluxes}
\label{sec:source_fluxes}

The count rates in three bands reported here were obtained with \texttt{CMLDetect}. Vignetting and quantum efficiency were taken into account when measuring the effective exposure time. The count rate error at 68\% confidence level was computed using the equation $err\_rate$=$\frac{\sqrt{C_{S,90\%}+(1+a)\times B_{90\%}}}{0.9\times T}$, where $C_S$ are the source net counts estimated by aperture photometry, using, for each observation where the source was detected, an extraction radius including 90\% of the EEF; $B$ are the background counts estimated in the same aperture on the background maps used in \texttt{CMLDetect} and corrected with a factor $a$=0.5, introduced to account for the uncertainties on the background estimation in a given position (see P09); $T$ is the vignetting corrected exposure time.

\begin{figure}[t]
\centering
\includegraphics[width=0.5\textwidth]{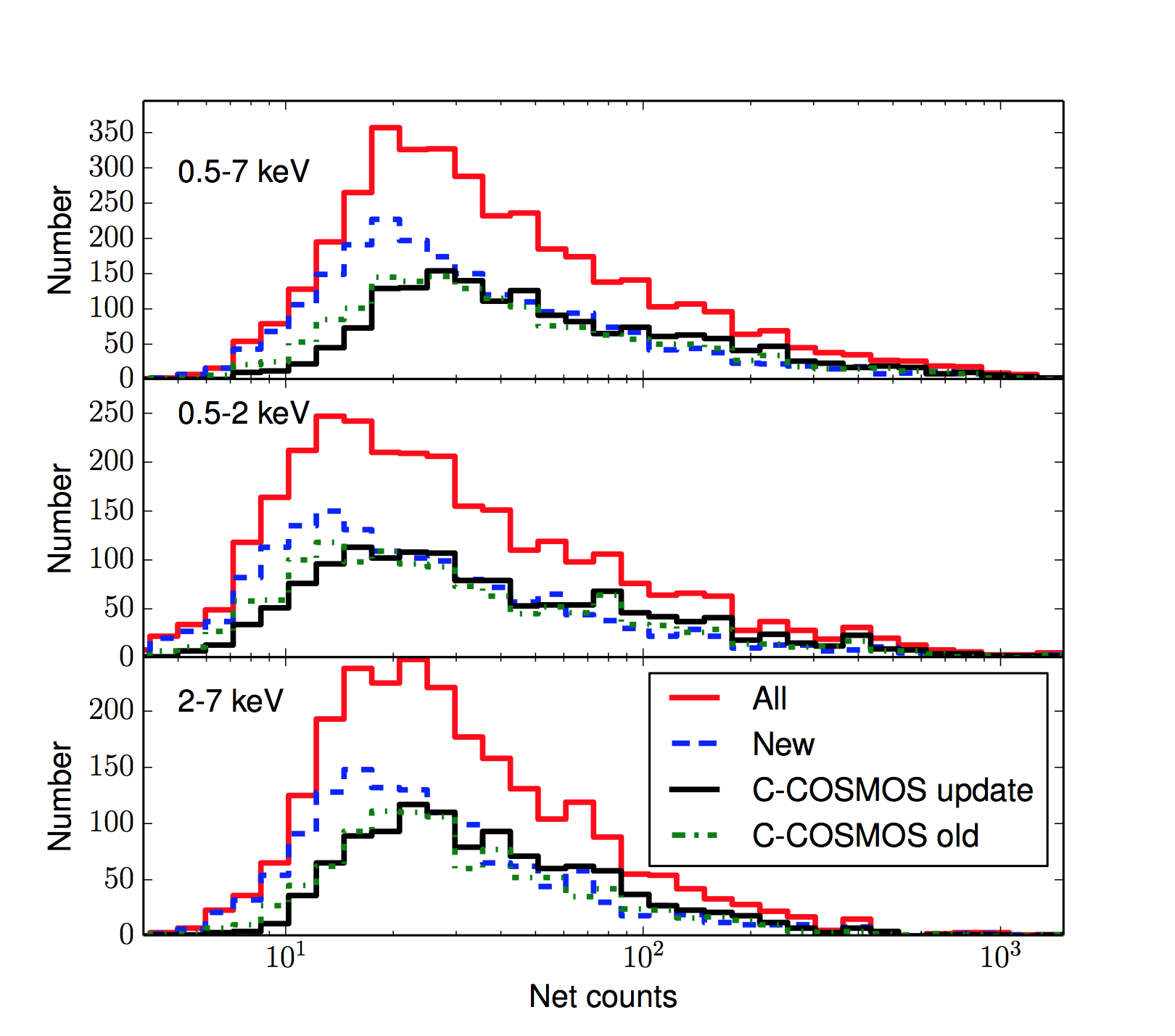}
\caption{Source count distributions in three bands: 0.5-7 keV (top), 0.5-2 keV (center) and 2-7 keV (bottom) for \leg\ (solid red), new data only (blue dashed) C-COSMOS old (green dot-dashed) and updated (black solid). Sources with upper limit have not been included. }
\label{counts_histo}
\end{figure}

In Figure \ref{counts_histo}, the net count distributions for the new sources in three bands are compared to those in E09 (C-COSMOS old) and also to the updated counts distribution of C-COSMOS (C-COSMOS new). The total is the sum of the new detections plus the updated C-COSMOS. 
The median (mean) value of net counts in the whole dataset in full, soft and hard bands is 30, 20 and 22 (80, 60, 43), respectively, compared to C-COSMOS where we had 33, 22 and 23 (88, 65, 46). The total number of net counts for the 676 C-COSMOS sources also detected in the new dataset is on average 60-80\% larger than the number of counts in C-COSMOS only. As a consequence, the updated C-COSMOS count histograms in Figure \ref{counts_histo} are all shifted to a higher numbers of counts. While in the full band the peak of the distribution is still around 30 counts, we more than double the number of sources with more than 70 full band counts, for which it is possible to perform individual X-ray spectral analysis, from 390 (Lanzuisi et al. 2013) to $\sim$950 sources in \leg. 

The fluxes were obtained from the count rates using the relation $F$=$R$$\times$(CF$\times$10$^{-11}$), where $R$ is the count rate in each band and CF is the energy conversion factor computed using the CIAO tool {\it srcflux}, assuming a power-law spectrum with slope $\Gamma$=1.4 and a Galactic column density $N_H$=2.6 $\times$10$^{20}$ cm$^{-2}$. Due to the fact that the observations have been taken in two different \chandra\ cycles, i.e. Cycle 8 for C-COSMOS and Cycle 14 for the new data, we used as CF a weighted mean of the factors in the different cycles, depending on the exposure time for each source accumulated in each cycle, to take into account for its variation ($\sim$15\% between the two cycles). The Cycle 14 (Cycle 8\footnote{The CF used for C-COSMOS and reported in E09 and P09 (computed using the online tool PIMMS) slightly differ from the one used here cause the latter are now computed with the most updated response matrix. The difference is $\sim$15\% and it reflects on the final fluxes for all C-COSMOS sources.}) CF are 1.71 (1.57), 7.40 (6.34) and 3.06 (3.04) counts erg$^{-1}$ cm$^{2}$ for 0.5-10, 0.5-2 and 2-10 bands, respectively. The conversion factors are sensitive to the assumed spectral shape: for $\Gamma$=2, there is a change of 40\% in the full band CF, of $\sim$5\% in the soft band and of $\sim$20\% in the hard band.

\begin{figure}[t]
\centering
\includegraphics[width=0.5\textwidth]{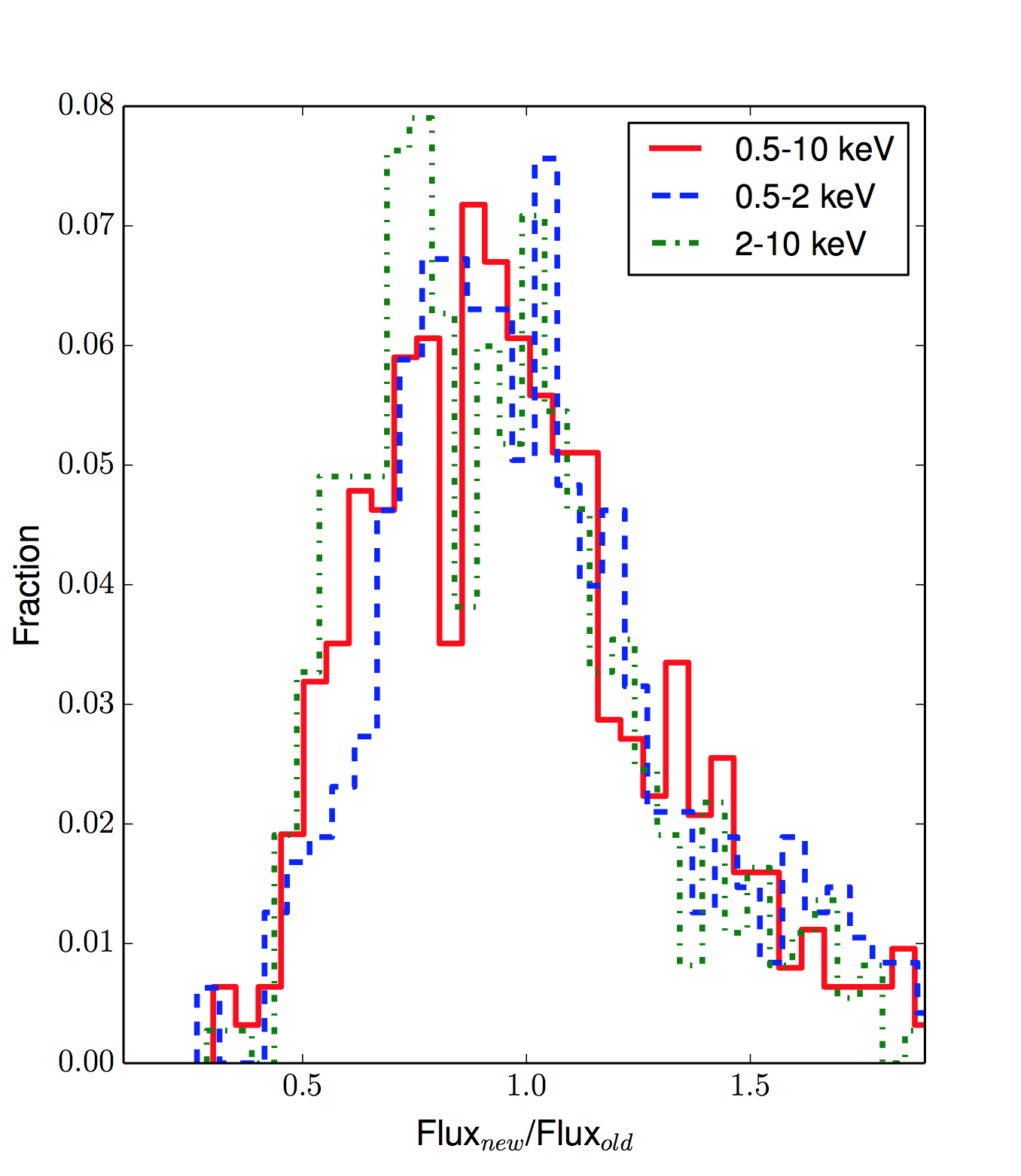}
\caption{Normalized distributions of ratios between new and old fluxes of the 676 sources detected in C-COSMOS and also in the new data at DET\_ML$>$10.8. Sources with upper limit have not been included.}
\label{fig:hist_flux_675}
\end{figure}

For the 676 C-COSMOS sources detected in the new data as well, we computed new total X-ray fluxes. In Figure \ref{fig:hist_flux_675} the normalized distribution of ratios between total and old fluxes are plotted for the three bands. From Gaussian fitting of the distributions, we find centroids at (F$_{new}$/F$_{old}$) =1.06, 1.11, 0.99 and standard deviations of $\sim$0.50, $\sim$0.55 and $\sim$0.40 in full, soft and hard band, respectively, showing a good agreement between old and new fluxes. The distributions show wings to both negative and positive values. Malmquist bias is most likely responsible for the negative wing, while variability for the positive one. 

\begin{figure}[t]
\centering
\includegraphics[width=0.5\textwidth]{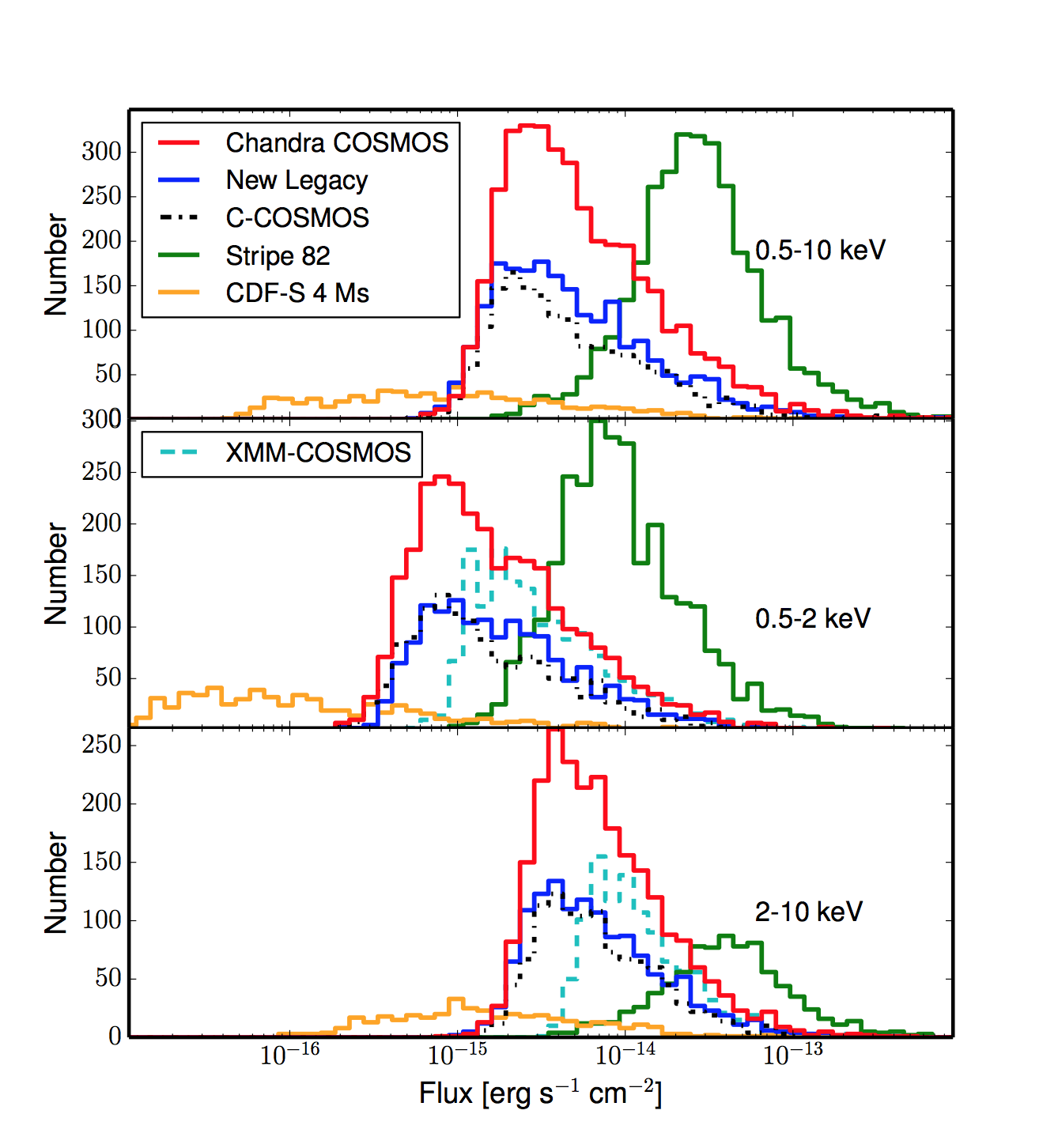}
\caption{Flux distributions for sources detected in 0.5-10 keV (top), 0.5-2 keV (center) and 2-10 keV (bottom) bands for \leg\ (solid red), new data only (blue solid), C-COSMOS updated (black dotted) and XMM-COSMOS (cyan dashed) sources. We also include the CDFS 4 Ms source flux distribution (orange) and the Stripe82 sources (green). Sources with upper limit have not been included.}
\label{fig:hist_flux}
\end{figure}

The distributions of X-ray fluxes for the whole \chandra\ COSMOS Legacy survey in the full, soft and hard bands is shown in Figure \ref{fig:hist_flux}, where it is also compared with C-COSMOS (the new version with just the updated fluxes, given the excellent agreement) and XMM-COSMOS. The new survey is about $\sim$2.5 times deeper than XMM-COSMOS in the 0.5-2 keV band and $\sim$2 times in the 2-10 keV band, and more than doubles the number of C-COSMOS sources in the same flux range. In the same Figure we compare our data with the 4 Ms CDFS (Xue et al. 2011)  and the large area Stripe82 survey (LaMassa et al. 2013a,b) source flux distributions, respectively to the left and to the right of \leg\ flux distribution. The combination of the three surveys (the deepest, the intermediate	and among the widest, see also Section \ref{summary}) allows to cover more than 4 orders of magnitude in flux.

Upper limits (90\% confidence level) on net counts, count rates and fluxes are given for all sources found in one band but not detected in another band. The upper limits were computed with the same procedure adopted for C-COSMOS and largely described in P09, to which we refer for a complete description.

\subsubsection{Hardness ratio analysis}
\label{hr}

In order to provide a rough estimate of the X-ray spectral shape of the sources, in particular of the intrinsic obscuration (see Marchesi et al. in press), for all the sources in the catalog, including the C-COSMOS sources, we computed the hardness ratio defined as $HR$=$\frac{H-S}{H+S}$, where $H$ are the net counts in the hard band and $S$ are those obtained in the soft band. Given the low number of counts for most of the sources (see Figure  \ref{counts_histo}), we used BEHR (Bayesian Estimation of Hardness Ratios, Park et al. 2006) which is particular effective in the low count regime, not needing a detection in both bands to work.

We extracted aperture photometry counts from each observation where the source was detected, using the PSF radius at encircled energy fraction (EEF)=0.9. We also extracted the background counts from the same observations, using an annulus with $r_{min}=r_{PSF}+8$ pixels and $r_{max}=r_{PSF}+40$ pixels, where $r_{PSF}$ is the PSF radius at encircled energy fraction (EEF)=0.95 (in pixel). In the background extraction, we excluded the contamination by other nearby detected sources using an exclusion radius equal to $r_{PSF}$. Total counts, background counts and the ratio between the sum of background areas and the sum of source areas, both in soft and hard bands, were then fed as input parameters to BEHR. 

\begin{figure}
\centering
\includegraphics[width=0.5\textwidth]{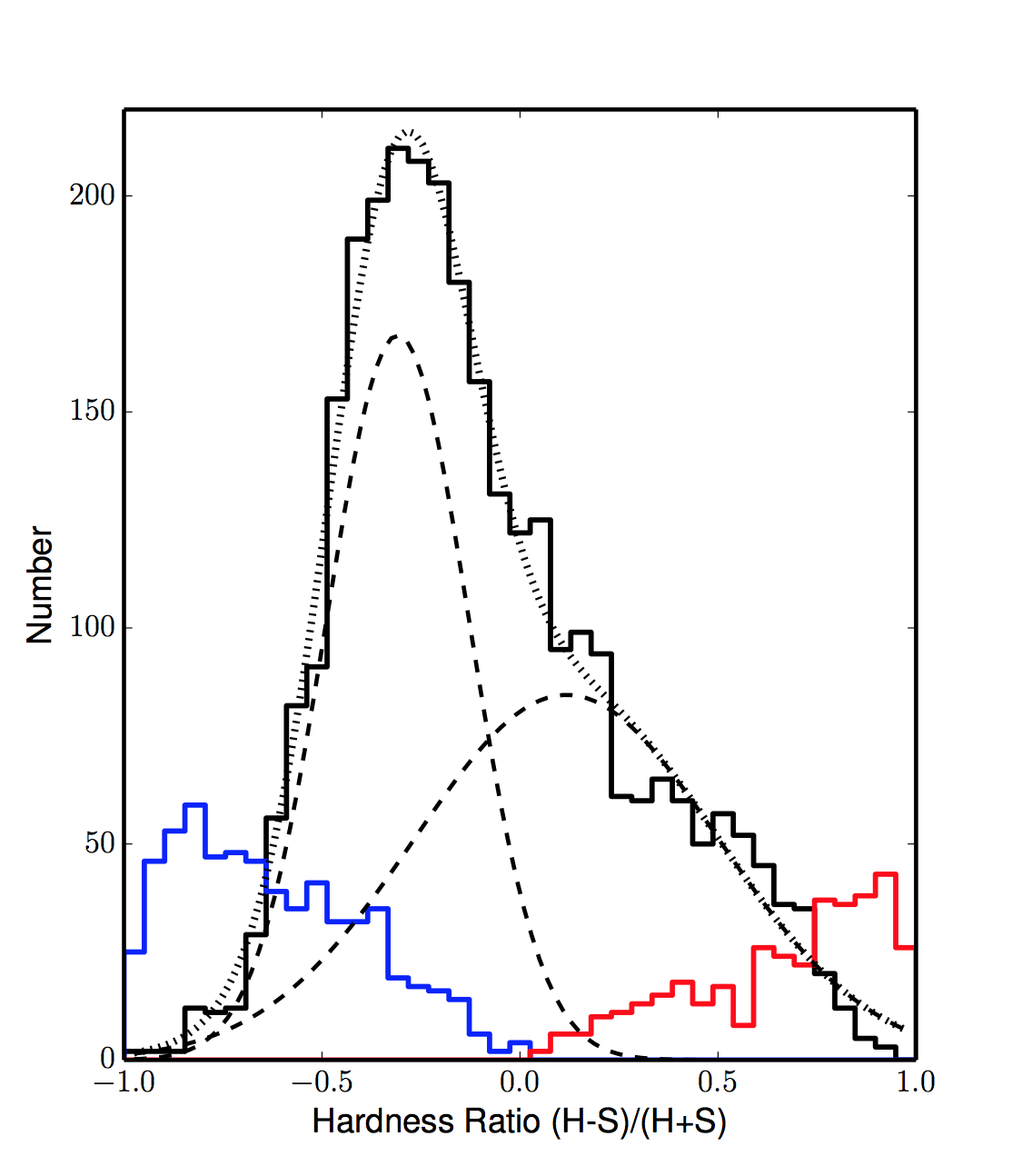}
\caption{HR distributions for the whole sample (black), upper limits (blue) and lower limits (red). The dotted line is the sum of the double Gaussian fitting. The dashed lines are the two Gaussian resulting from the fitting.}
\label{hr_dist}
\end{figure}

For most sources ($>$3000) BEHR finds a detection on the HR, and for 989 sources an upper or lower limit (616 and 371 sources respectively). The typical error on the HR is $\sim$0.2. In Figure \ref{hr_dist}, we plot the distribution of the HRs for the measured values (black solid line), for the lower limits (red) and the upper limits (blue). The mean (median) HR value is -0.09 (-0.17) for the measured values and it moves to lower values when including upper and lower limits (-0.11 and -0.19 for the mean and the median, respectively). A Gaussian fit returns a peak at -0.20 with a 1$\sigma$ dispersion of 0.32, however a single Gaussian is not clearly a best representation of the HR distribution. A double Gaussian fit returns a peak at -0.31 and one at 0.12 with a 1$\sigma$ dispersion of 0.18 and 0.38, respectively. 

Hardness ratio is not a fully reliable measurement of obscuration, because of the complexity of the spectral shape, the large error bars due to low counts statistic and the redshift dependency (see Marchesi et al. in press), however it is possible to roughly assume an HR value to divide the sources in obscured and unobscured. 
We use here HR=--0.2, which has been shown to be a fair value to separate sources with column densities above and below 10$^{22}$ cm$^{-2}$ (Lanzuisi et al. 2013, Civano et al. 2012) at all redshifts. A total of 1993 sources, 50$^{+17}_{-16}$\% of the entire sample (errors have been computed using HR 1$\sigma$ errors) are therefore classified as obscured. Tentatively, the double Gaussian fit of the HR distribution could also be interpreted as to be due from two populations of sources, the obscured population peaking at positive HRs and the unobscured population peaking at negative HR. The broad dispersion of the Gaussian peaking at positive HR could be due to high redshift obscured sources whose HR would be negative even if obscured. A more detailed analysis on the obscured AGN fraction is presented in Marchesi et al. (in press).

\subsubsection{Source catalog}

The catalog released with this paper contains all the measurements discussed above. In Table \ref{tab:cat_summary}, we show the columns of the catalog of the new 2273 sources (named as
``lid'' in column 1) combined with the updated C-COSMOS catalog of 1743 sources (named as ``cid'' in column 1). The catalog will also be stored in the COSMOS web site at the \leg\ project page\footnote{http://irsa.ipac.caltech.edu/data/COSMOS/tables/chandra/}. Data products, including exposure and events mosaics, are available in the dedicated page\footnote{http://irsa.ipac.caltech.edu/data/COSMOS/images/chandra/} at the same website.

\begin{table}[t]
\centering
\begin{tabular}{cccc}
\hline
Bands & New & C-COSMOS & Legacy\\
\hline
F+S+H & 1140 & 1047 (922) & 2187\\
F+S & 536 & 397 (474) & 933\\
F+H & 448 & 231 (257) & 679\\
F & 121 & 49 (73) & 170\\
S & 21 & 17 (32) & 38\\
H & 7 & 2 (3) & 9\\

\textbf{Total} & \textbf{2273} & \textbf{1743 (1761)} & \textbf{4016}\\
\hline
\end{tabular}
\caption{Number of sources with  DET\_ML$>$10.8 in at least one band, for each combination of X-ray bands. The ``New" column includes all the new detected sources. The columns labelled as ``C-COSMOS'' include the updated numbers, using the information from the new data and in parenthesis also the old numbers as in Elvis et al. (2009). }\label{tab:x_data}
\end{table}

\begin{table}[t]
\centering
\begin{tabular}{c|cc|cc|cc}
\hline
Bands & \multicolumn{2}{|c|}{New} &  \multicolumn{2}{|c|}{C-COSMOS} &  \multicolumn{2}{|c}{Legacy}\\
& $>$ 7& $>$12 & $>$ 7& $>$12 & $>$ 7& $>$12\\
\hline
F & 5 &5 & 6 &6 & 12 &11 \\
S & 4 &3 &4&3 & 9&7\\
H& 3 & 3&4&3& 8&7\\
\hline
\end{tabular}
\caption{Number of spurious sources with DET\_ML$>$10.8 with at least 12 and 7 full band counts, corresponding to a reliability of 99.7\% for the new data, the old C-COSMOS data (as in P09 Section5) and in the whole \leg. }\label{tab:spurious}
\end{table}

\begin{table*}[t]
\centering
\begin{tabular}{cccc|ccc}
\hline
Band & &  DET\_ML$\geq$10.8 & & & 6$<$ DET\_ML$<$10.8\\
& New & C-COSMOS & Legacy & New & C-COSMOS & Legacy\\
\hline
Full (F) & 2146 & 1667 (1655) & 3813 & 99 & 57 (71) & 156\\
Soft (S) & 1538 & 1382 (1340) & 2920 & 159 & 79 (88) & 238 \\
Hard (H) & 1325 & 1115 (1017) & 2440 & 271 & 165 (165) & 436\\
\hline
\end{tabular}
\caption{Number of sources detected in each band at the two adopted thresholds. The columns labelled as C-COSMOS include the updated numbers using the information from the new data and, in parenthesis, also the old numbers as in Elvis et al. (2009).}\label{tab:x_single}
\end{table*}

\begin{table*}
\centering
\scalebox{0.7}{
\begin{tabular}{ccc}
\hline
\hline
No. & Field & Note\\
\hline
1 & Name & \chandra\ source name\\
2 & R.A. & \chandra\ Right Ascension (J2000, hms)\\
3 & DEC & \chandra\ Declination (J2000, dms)\\
4 & pos\_err & Positional error [arcsec]\\
 \hline
5  &  DET\_ML\_F & maximum likelihood detection value in 0.5-7 keV band \\
6 & rate\_F & 0.5-7 keV count rate [counts s$^{-1}$]\\
7 & rate\_F\_err & 0.5-7 keV count rate error [counts s$^{-1}$]\\
8 & flux\_F & 0.5-10 keV flux [erg cm$^{-2}$ s$^{-1}$]\\
9  & flux\_F\_err & 0.5-10 keV flux error [erg cm$^{-2}$ s$^{-1}$]\\
10 & snr\_F & 0.5-7 keV S/N Ratio\\
11 & exptime\_F & 0.5-7 keV exposure time [ks]\\
12 & cts\_ap\_F & 0.5-7 aperture photometry counts [counts]\\
13 & cts\_ap\_F\_err & 0.5-7 aperture photometry counts error [counts]\\
\hline
14   &  DET\_ML\_S & maximum likelihood detection value in 0.5-2 keV band \\
15 & rate\_S & 0.5-2 keV count rate [counts s$^{-1}$]\\
16 & rate\_S\_err & 0.5-2 keV count rate error [counts s$^{-1}$]\\
17 & flux\_S & 0.5-2 keV flux [erg cm$^{-2}$ s$^{-1}$]\\
18 & flux\_S\_err & 0.5-2 keV flux error [erg cm$^{-2}$ s$^{-1}$]\\
19 & snr\_S & 0.5-2 keV S/N Ratio\\
20 & exptime\_S & 0.5-2 keV exposure time [ks]\\
21 & cts\_ap\_S & 0.5-2 aperture photometry counts [counts]\\
22 & cts\_ap\_S\_err & 0.5-2 aperture photometry counts error [counts]\\
 \hline
23   &  DET\_ML\_H & maximum likelihood detection value in 2-7 keV band \\
24 & rate\_H & 2-7 keV count rate [counts s$^{-1}$]\\
25 & rate\_H\_err & 2-7 keV count rate error [counts s$^{-1}$]\\
26 & flux\_H & 2-10 keV flux [erg cm$^{-2}$ s$^{-1}$]\\
27 & flux\_H\_err & 2-10 keV flux error [erg cm$^{-2}$ s$^{-1}$]\\
28 & snr\_H & 2-7 keV S/N Ratio\\
29 & exptime\_H & 2-7 keV exposure time [ks]\\
30 & cts\_ap\_H & 2-7 aperture photometry counts [counts]\\
31 & cts\_ap\_H\_err & 2-7 aperture photometry counts error [counts]\\
 \hline
32   & hr & Hardness ratio\\
33 & hr\_lo\_lim & Hardness ratio 90\% lower limit\\
34 & hr\_up\_lim & Hardness ratio 90\% upper limit\\
\hline
\hline

\end{tabular}}\caption{Data fields in the catalog.}\label{tab:cat_summary}
\end{table*}

\subsection{Matching with XMM-COSMOS catalog}

We matched the \leg\  sources with those in XMM-COSMOS (Cappelluti et al. 2009). There are 1714 secure XMM-COSMOS sources with at least one counterpart in \leg, 824 of which have at least one counterpart in the new data.  There are 46 XMM-COSMOS sources outside the area covered by \leg\ (see Fig. \ref{tiling}) and 126 with no \chandra\ counterparts.
In summary, 93\% of the XMM-COSMOS sources within the \leg\ area have at least one \chandra\ counterpart. 

The 126 sources with no \chandra\ counterparts can be divided in three groups: the 25 sources (20\%) with \chandra\ exposure $<$40 ks; the 60 sources (48\%; 13 of these sources have also \chandra\ exposure $<$40 ks) with XMM-COSMOS DET\_ML$<$15 in all of the three bands (0.5-2 keV, 2-8 keV, 4.5-8 keV); last, the 54 sources with XMM-COSMOS DET\_ML$>$15 in at least one band and \chandra\ exposure $>$40 ks. For the first group, the low exposure time could be the reason of the non detection, while for the second a non-detection in \chandra\ can be explained with a flux fluctuation within the flux uncertainty. We visually inspected the sources in the last group and we found that seven of them are located inside a bright cluster, and therefore have been not resolved into point sources by our analysis. For the remaining 47 sources the \chandra\ signal is weak or negligible, and therefore these sources could be candidate variable AGN. In particular, XMM-ID 30748 has DET\_ML 20 times larger than the detection threshold in XMM-COSMOS: this source was detected only in the 0.5-2 keV band, with a flux $F$=2.7$\times$10$^{-15}$ erg s$^{-1}$ cm$^{-2}$ and a photometric redshift $z$=2.71. Despite being interesting and worth further analysis on the variability, this is beyond the scope of this paper. 

There are 58 XMM-COSMOS sources that have been resolved by the smaller \chandra\ PSF into two distinct sources using a maximum radius of 10\arcsec\ for the match. Two XMM-COSMOS sources have been resolved into three \chandra\ sources using a maximum radius of 10\arcsec. As a comparison, 25 XMM-COSMOS sources (Brusa et al. 2010) were resolved into two separate C-COSMOS sources. More details on the optical counterparts of the XMM-COSMOS sources resolved in two \chandra\ ones are given in Marchesi et al. (in press).

There is a good agreement between XMM-COSMOS and \chandra\ fluxes. We rescaled the \chandra\ \leg\ fluxes using the same slope used for XMM-COSMOS ($\Gamma$=2 in soft band and $\Gamma$=1.7 in hard band) and we found that the median value of the ratio $flux_{XMM} / flux_{\chandra}$ is 1.13 in soft band and 1.22 in hard band. 

\section{Sky coverage and survey sensitivity}
\label{sec:sensitivity}

The sky coverage of a survey is the area covered as a function of the flux limit. We computed it in three bands (0.5-10, 0.5-2 and 2-10 keV) using the exposure and background maps (see Section \ref{sec:expo} and \ref{sec:bkg}) produced for the source detection, and assuming a power-law spectrum with $\Gamma$=1.4 and Galactic N$_H$=2.6$\times$10$^{20}$ cm$^{-2}$. X-ray observations have a flux limit that changes over the field of view because the \chandra\ PSF changes in both size and shape as a function of the distance from the aim point and because the effective area is vignetted. In this survey, where the total coverage is obtained using multiple overlapping pointings, every source was observed in up to six different positions on the detector, resulting in a quite uniform average PSF (Figure \ref{psf_size}). 

The procedure we used to compute \leg\ survey sky coverage is closely similar to that used by P09 for C-COSMOS, but makes use of a PSF map for each observation instead of an analytical form of the PSF as function of the off-axis angle. This is a more time consuming approach but one that returns a more detailed sensitivity map, which can be valuable in other studies (e.g., clustering analysis and correlation functions) or simply for source photometry (Section \ref{hr}). 

For each observation we made use of the CIAO tools \texttt{mkpsfmap} and \texttt{dmimgadapt} to create a background map convolved with the PSF map in such a way that at each position of the map, the count value corresponds to the number of counts in an aperture corresponding to 50\% of the encircled energy fraction at that position. 

For each position of the entire mosaic (applying a binning of 8 pixels for computing time purposes), we computed the minimum number of counts $C_{min}$ needed to exceed the background fluctuations, assuming the same probability for spurious sources (i.e., DET\_ML threshold) used in the C-COSMOS and \leg\ catalogs for the Poisson statistics, i.e. 2$\times$10$^{-5}$. We used the relation 
\begin{equation}
\label{eq:prob}
P_{Poisson}=e^{-B} \sum_{k=C_{min}}^\infty \frac{B^k}{k!} = 2\times10^{-5},
\end{equation}

where $B$ is the total background counts computed at each position of the grid, by summing the background counts in each observation covering that given position. Equation \ref{eq:prob} is solved iteratively to find $C_{min}$; then the count rate limit, $R_{lim}$ is obtained using

\begin{equation}
R_{lim}=\frac{C_{min}-B}{f_{psf}\times T_{exp}},
\end{equation}

where $T_{exp}$ is the total, vignetting corrected, exposure time at each position on the grid, while $f_{psf}$ is the encircled count fraction of the PSF. In C-COSMOS, this value was tuned to reproduce the simulation results and then it was fixed to $f_{psf}$=0.5, however any number in the range 0.5-0.9 produced similar results with variations of the order of few percent in the resulting sensitivity.

Finally, we converted the count rate limit $R_{lim}$ into the flux limit using the same conversion factors used for the sources in the catalog based on the position (see Section \ref{sec:source_fluxes}). We also computed the sensitivity for only the C-COSMOS area with the same method and obtained the same sensitivity as published in E09 and P09.

\begin{figure}
\centering
\includegraphics[width=0.5\textwidth]{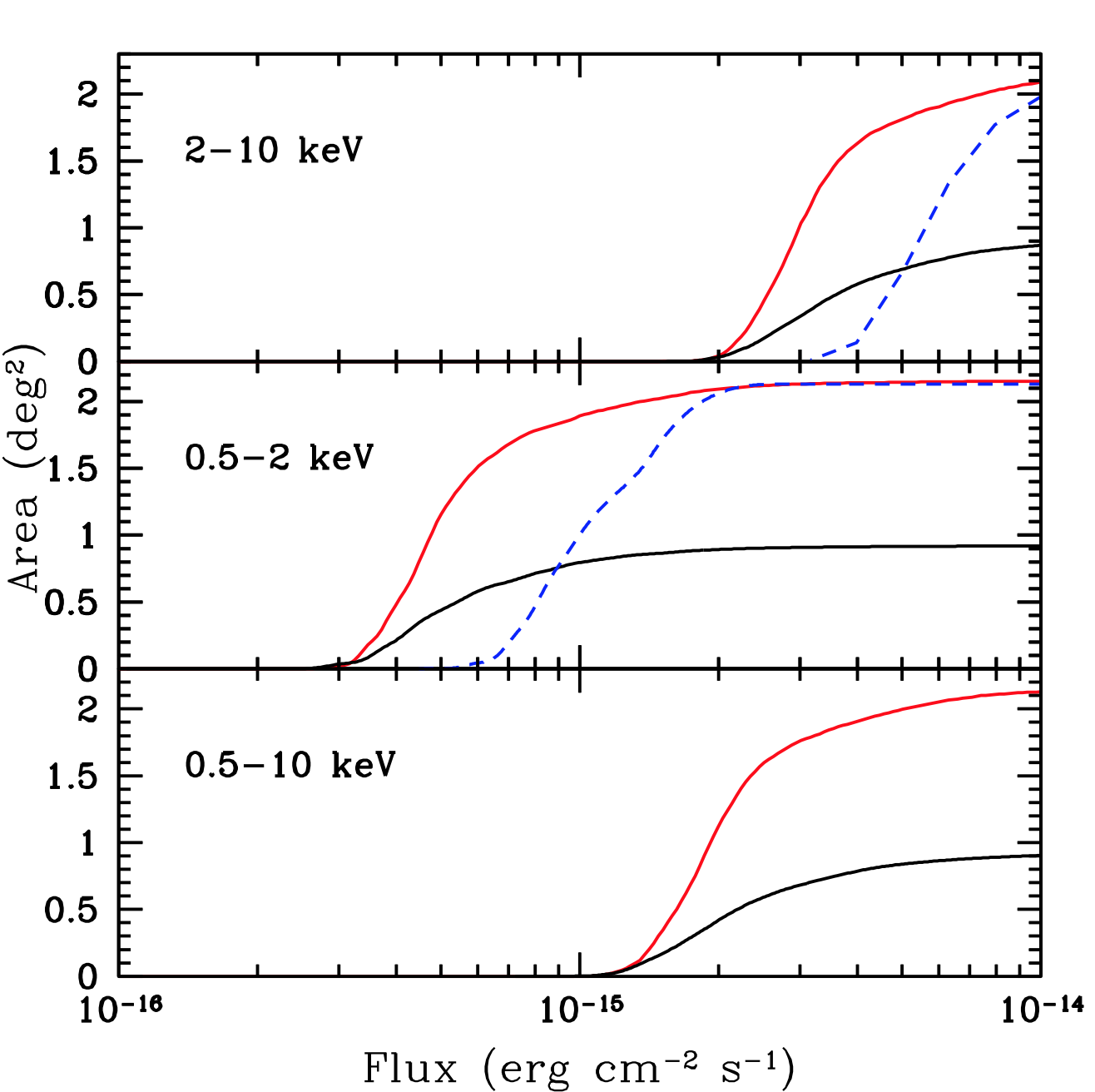}
\caption{Area-flux curve for \leg\ (red solid line) in 2-10 keV (top), 0.5-2 keV (center) and 0.5-10 keV (bottom) bands. The coverage of C-COSMOS (black solid line) and XMM-COSMOS in the 0.5-2 keV and 2-10 keV bands (Cappelluti et al. 2009; dashed blue line) are shown for comparison. }
\label{fig:sens}
\end{figure}

The sky coverage of the \chandra\ COSMOS Legacy survey in the three energy bands is shown in Figure \ref{fig:sens}. We compare our results with those of C-COSMOS (black solid lines) and XMM-COSMOS (blue dashed lines): the new survey covers a similar area to XMM-COSMOS and almost three times the area of C-COSMOS at faint fluxes (e.g., $\sim 5\times 10^{-15}$ \cgs\ in the soft band) and $\sim$2 times at bright fluxes (e.g., $> 10^{-15}$ \cgs\ in the soft band).

 We have verified that the limits at 20\% (50\%) completeness for the Legacy catalog are consistent with those computed and reported in Table 2 of P09 and assuming the changes in CF used here and explained in Section \ref{sec:source_fluxes} of 1.5 (1.9) $\times10^{-15}$, 3.9 (4.9)$\times10^{-16}$ and 2.5 (3.1) $\times10^{-15}$ \cgs\ in the F, S, and H bands. At this limit, \leg\ increases by a factor of 3 the area covered with respect to C-COSMOS.

\section{Number counts}
\label{sec:lgnlgs}

The log$N$-log$S$ relation, i.e. the number of sources $N(>S)$ per square degree detected at fluxes brighter than a given flux S (erg s$^{-1}$cm$^{-2}$), provides a first estimate of source space density as a function of flux and therefore information on the cosmic population to compare with different models of population synthesis. Given that multiple log$N$-log$S$ curves have been published in the literature, it is also a standard check to validate the many calibration steps used to produce a catalog of X-ray point like sources. 

We constructed the log$N$-log$S$ curve for \leg\ in both the 0.5-2 keV and 2-10 keV bands. Following P09, we included only sources with DET\_ML$>$10.8 and we applied a cut in SNR ($>2$ and $>$ 2.5 in soft and hard) to limit the Eddington bias effect, which could have a significant (up to 30-50\%) contribution at the lowest fluxes. This choice avoids sources with large statistical uncertainties on their fluxes and limits the errors due to the sky coverage uncertainties at the faint end. With the adopted thresholds in SNR, the agreement measured in P09 between simulations input and output log$N$-log$S$ is better than 5\%. {The procedure used by P09 is consistent with the one applied by Luo et al. (2008) on \chandra\ Deep Field South data. } The number of sources not included because of the SNR cut is $\sim$1\% in the soft and $\sim$5\% in the hard band.

The adopted SNRs imply the following flux limits: 2.7$\times$ 10$^{-16}$ erg s$^{-1}$ cm$^{-2}$ in the 0.5-2 keV band and 1.8$\times$ 10$^{-15}$ erg s$^{-1}$ cm$^{-2}$ in the 2-10 keV band. These are the same flux limits of C-COSMOS, which is expected given that the new observations have the same maximum exposure. The final number of sources used here for the number counts with the above constraints are 2758 in the soft band (1309 from C-COSMOS and 1449 from the new sample) and 2243 in the hard band\footnote{We also applied a cut in exposure time at 40 ks in the hard band to limit sources (65 in total) at the edges of the field with high background level.} (1056 from C-COSMOS and 1187 from the new sample).

\begin{figure*}
  \includegraphics[width=0.5\textwidth]{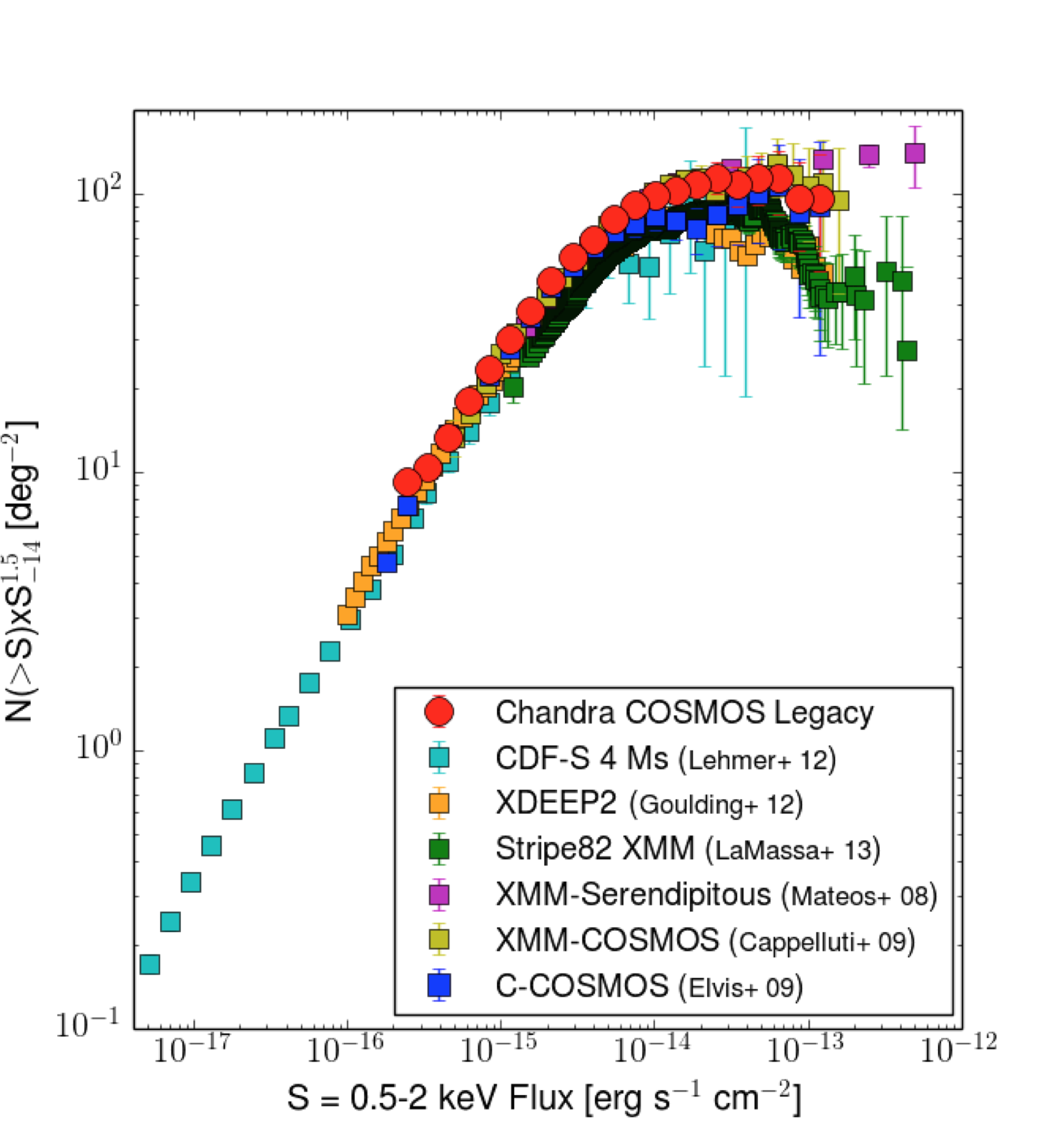}
  \includegraphics[width=0.5\textwidth]{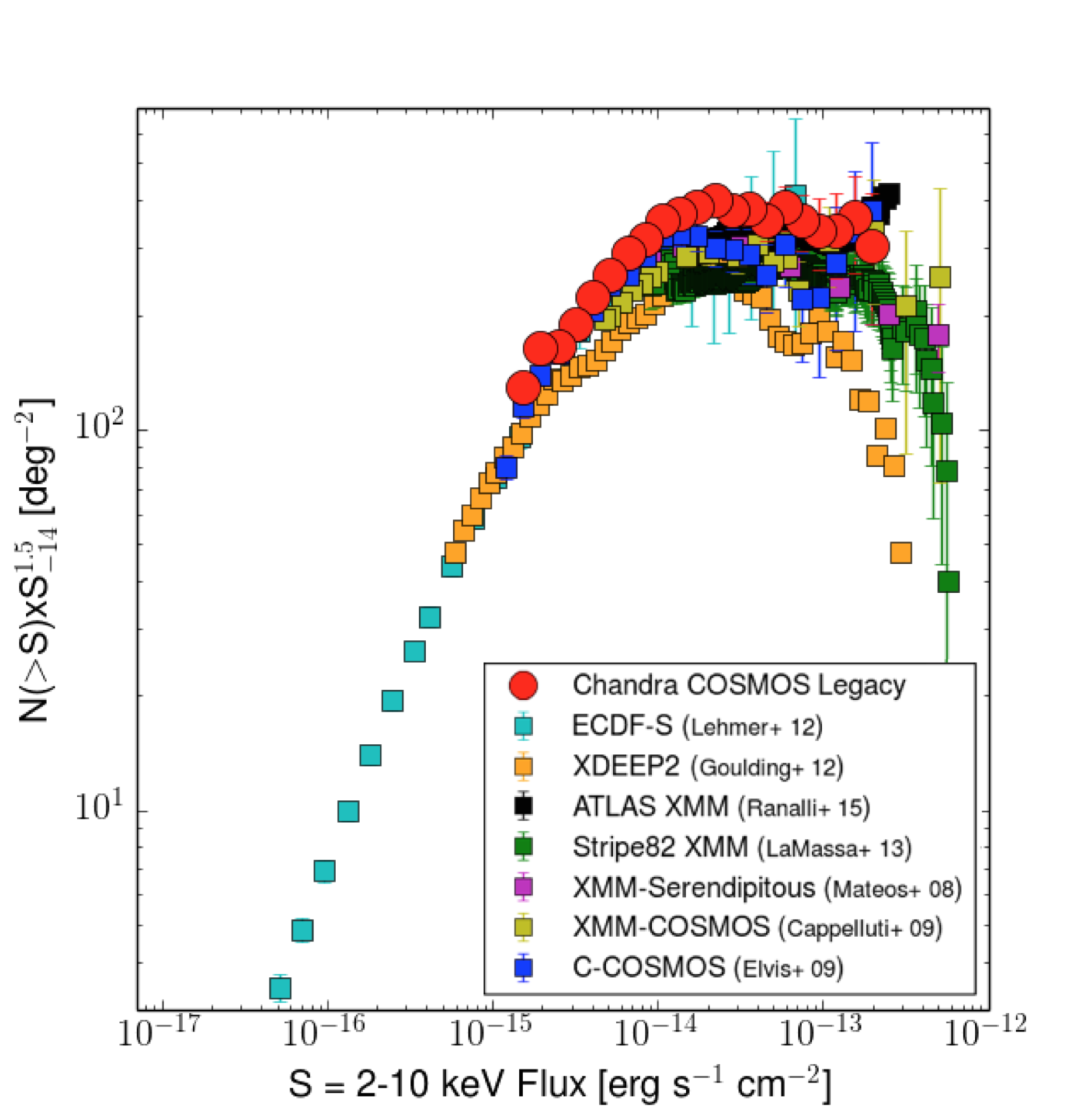}
  \includegraphics[width=0.5\textwidth]{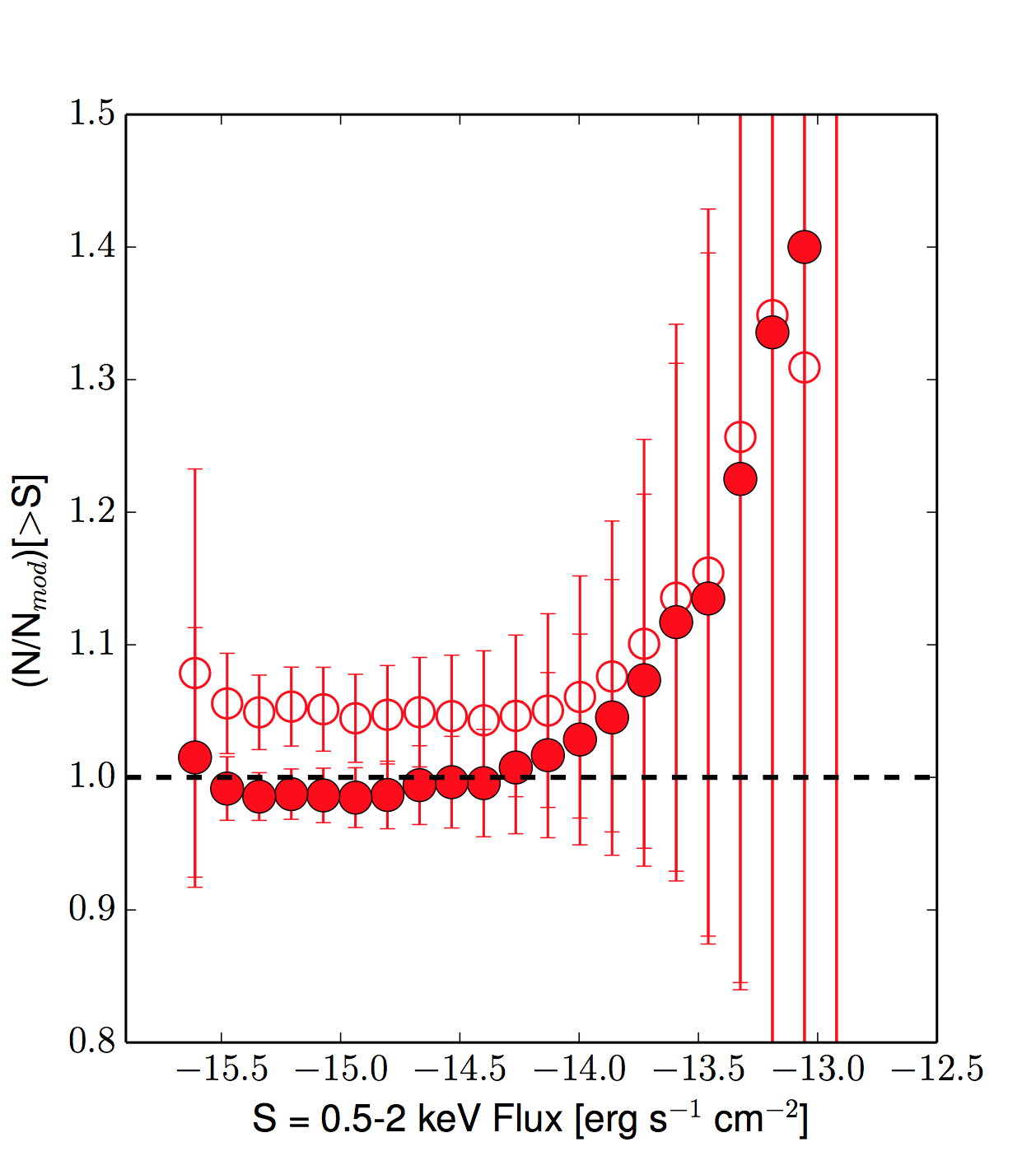}
  \includegraphics[width=0.5\textwidth]{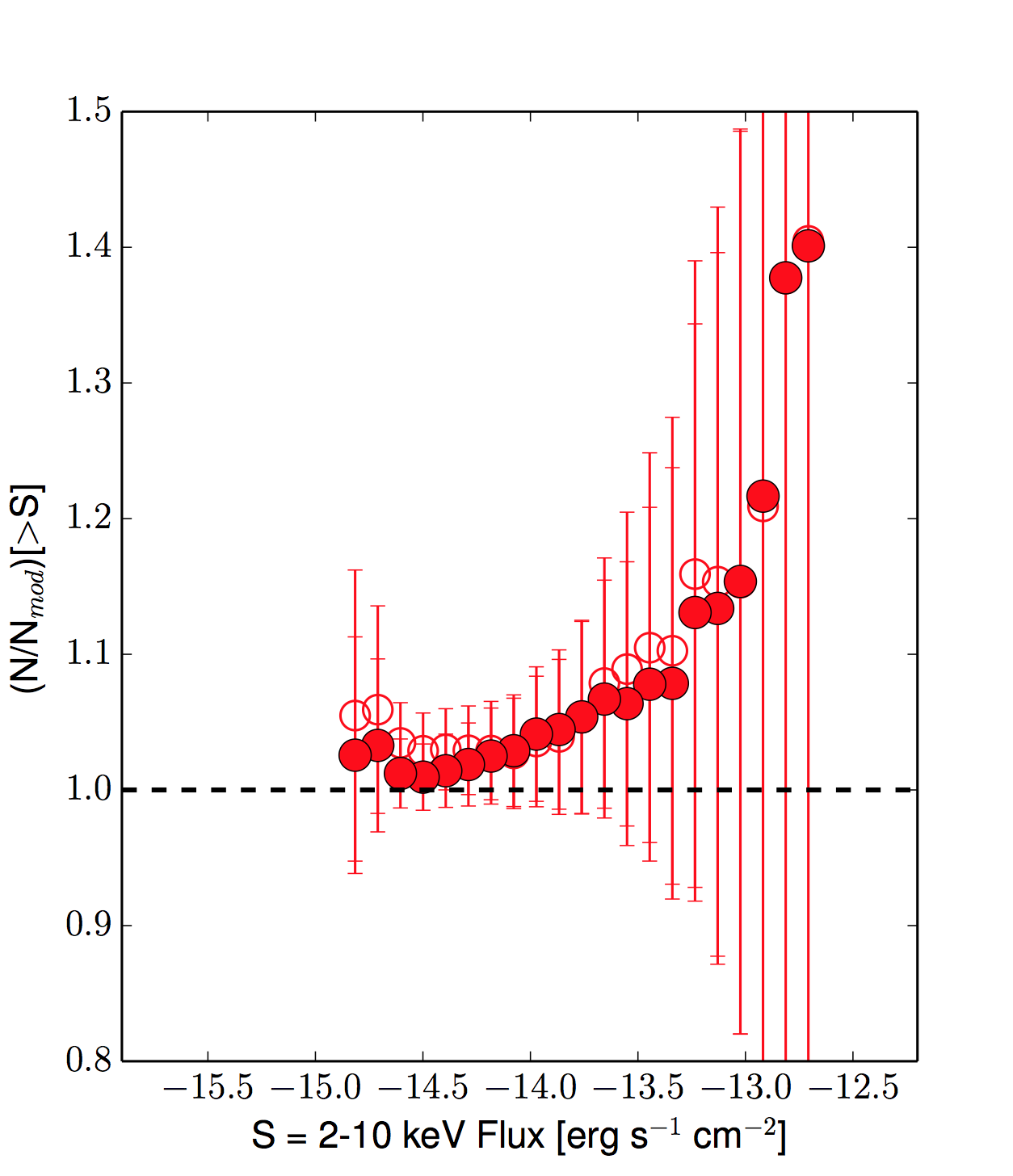}
  
\caption{Euclidean normalized log$N$-log$S$ curves in 0.5-2 keV (top-left) and 2-10 keV (top-right) bands. The \leg\ curve for all sources with  DET\_ML$>$10.8 and SNR$>$SNR$_{lim}$ is plotted in red circles. Results from previous works are plotted (see label in the plot). The ratio of \leg\ number counts to Gilli et al. (2007, red solid) and Treister et al. (2009; red empty) models are plotted in the soft and hard bands (bottom left and bottom right). The source number counts are multiplied by $(S /10^{14})^{1.5}$ to highlight the deviations from the Euclidean behavior. }
\label{fig:logn-logs}
\end{figure*}

We show the results obtained with these source selections in Figure \ref{fig:logn-logs} (top panels): the normalized Euclidean curves, i.e. with $N(>S)$ multiplied by $S^{1.5}$, are presented in order to enhance the differences between different surveys. In the same figure we include the C-COSMOS (E09) and XMM-COSMOS points (Cappelluti et al. 2009). 
We also compare our log$N$-log$S$ relationships with those from previous X-ray surveys, spanning from wide (Stripe82 XMM: LaMassa et al. 2013b; 2XMMi: Mateos et al. 2008), to moderate (XDEEP2: Goulding et al. 2012), to small areas (4 Ms CDFS: Lehmer et al. 2012). As XDEEP2 and CDFS define their hard band in a slightly different energy range, we converted their energy to 2-10 keV to perform an adequate comparison. 

\leg\ log$N$-log$S$ covers 3 and 2.5 orders of magnitude in flux in the soft and hard band, respectively, with 2-8\% errors at fluxes $<$$1-3\times$10$^{-14}$ \cgs, respectively. The excellent statistics allows to considerably reduce the uncertainties (20-30\%) in the number counts also at bright fluxes, which are now $\simeq$40\% smaller than in C-COSMOS.
 
In the soft band, there is an excellent agreement between our survey and previous works below $S$$\sim$10$^{-14}$ \cgs. At brighter fluxes instead, the uncertainties are larger due to the low number of detections (65 sources in \leg). A larger spread is observed when comparing results from different surveys, due to the fact that bright sources can be properly sampled only with extremely large areas ($>$5-10 deg$^2$). In the hard band instead, \leg\ number counts agree with other surveys at faint fluxes, while at the bright end (i.e. $S$$>$2$\times$10$^{-14}$ \cgs), the \leg\ counts are in the upper envelope of the spread.

We also compare our results with predictions of two different phenomenological models, Gilli et al. (2007) and Treister et al. (2009), assuming column densities in the interval N$_H$=10$^{20-26} cm^{-2}$ and redshift $z$=0-6. In Fig. \ref{fig:logn-logs} (bottom panels), we show the ratio of \leg\ number counts to both models in the soft and hard bands (left and right). At the faint end of the soft band, i.e., up to fluxes $\sim$10$^{-14}$ \cgs, our results are in agreement with the Gilli et al. (2007, solid points) model prediction within 1-5\%, while the Treister et al. (2009, open points) model (open points) slightly under-predictions the counts by 5--10\% in the same flux range.  At bright fluxes, where the sample is limited by the statistics, the differences between models and data becomes larger, even exceeding 10\%. 
In the hard band, both models reproduce well the observed data within 5\% below $>$2$\times$10$^{-14}$ \cgs\ and the difference becomes more pronounced at bright fluxes ($>$10\% at fluxes $>$5$\times$10$^{-14}$ \cgs)

\begin{figure*}
  \includegraphics[width=0.5\textwidth]{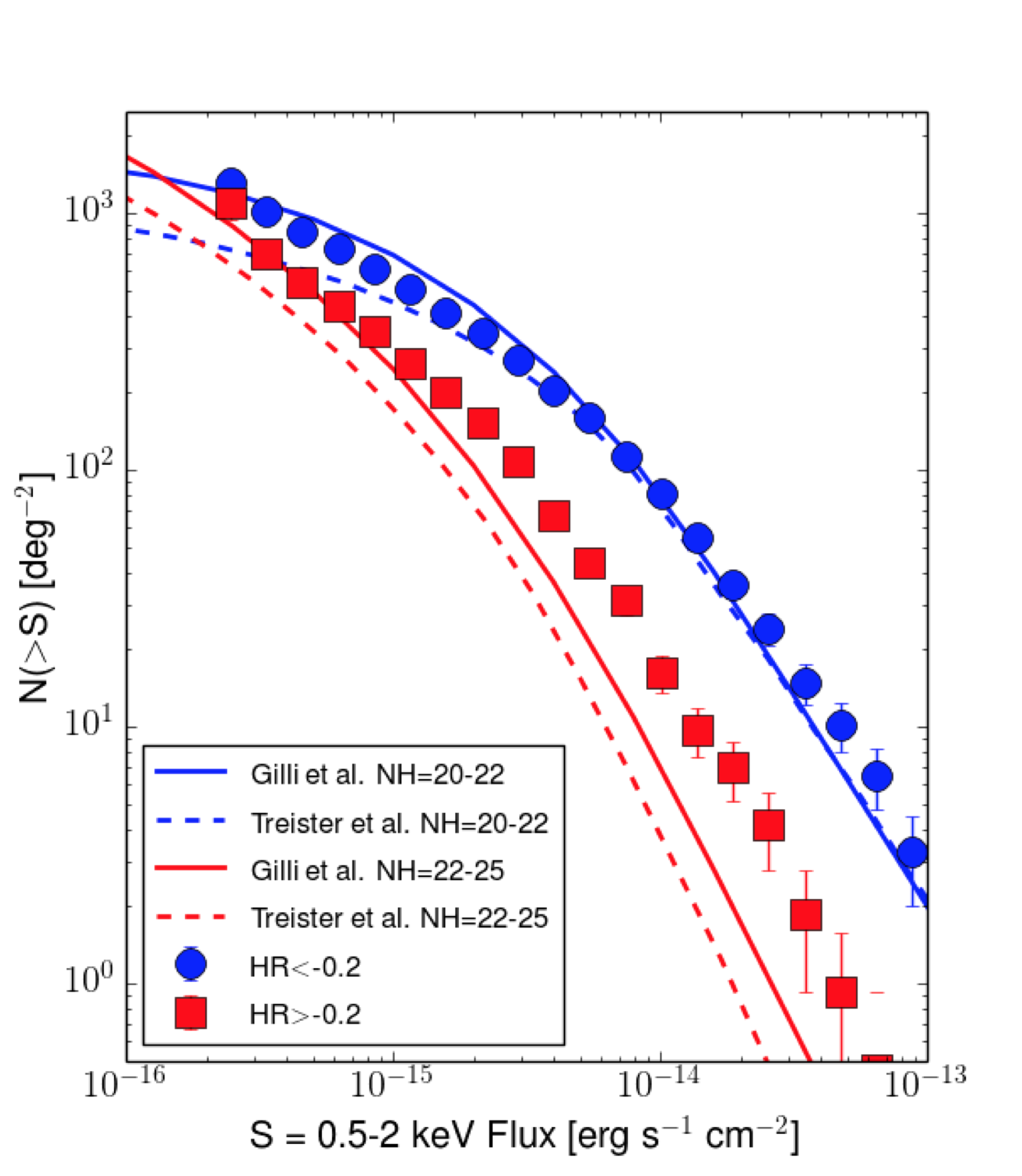}
  \includegraphics[width=0.5\textwidth]{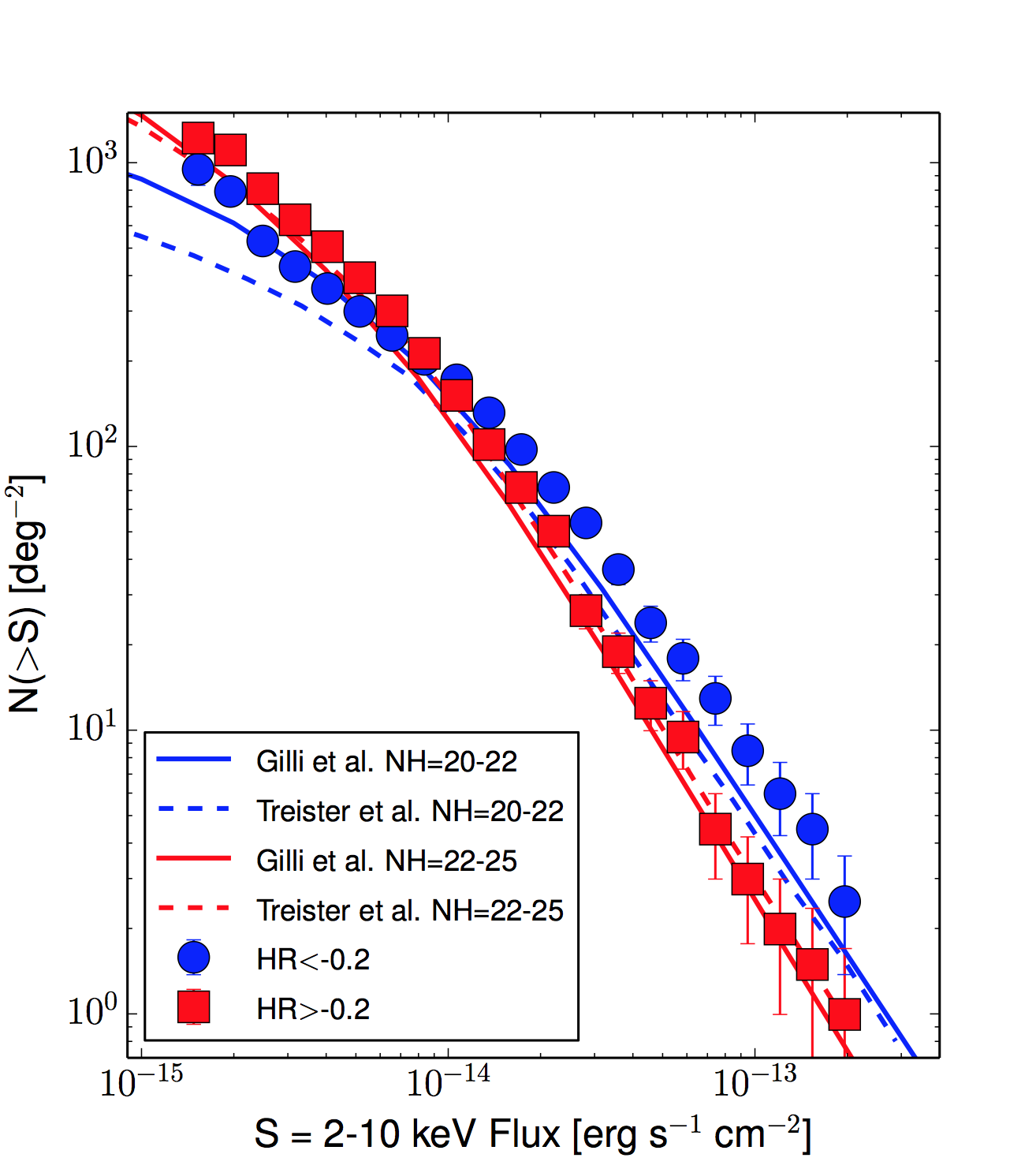}
 \includegraphics[width=0.5\textwidth]{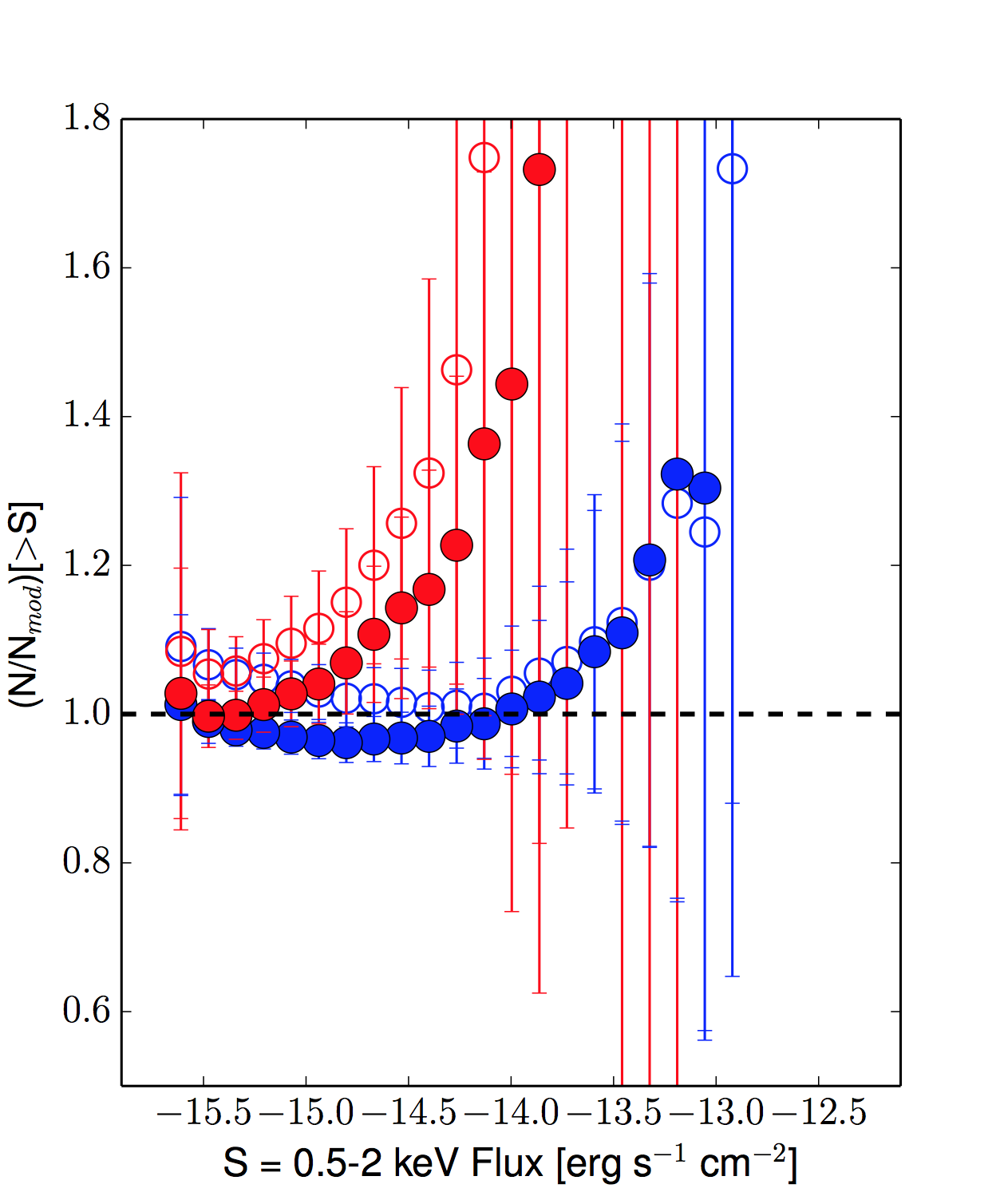}
  \includegraphics[width=0.5\textwidth]{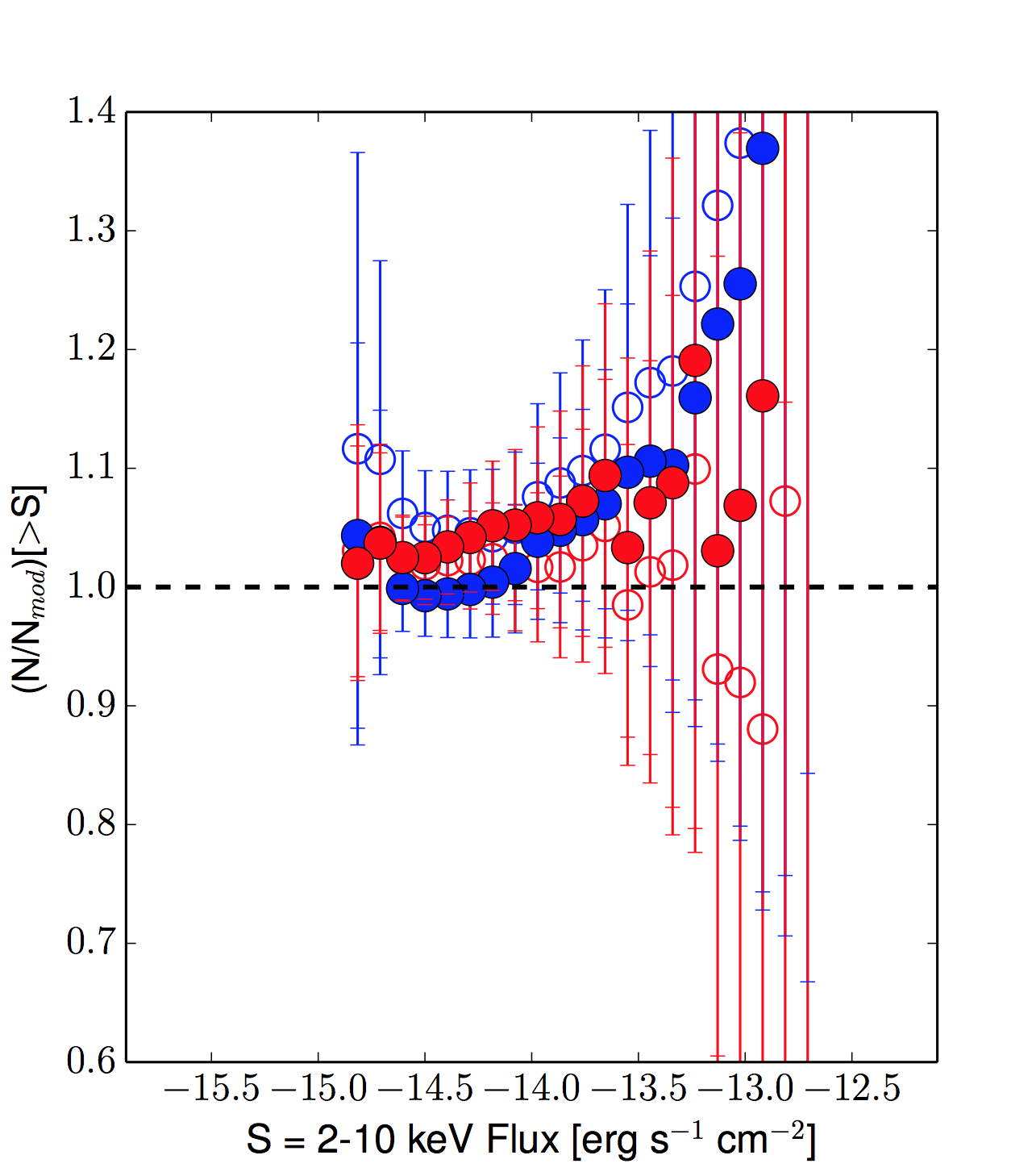}
\caption{Number counts in soft (top-left) and hard (top-right) bands for sources with HR$>$-0.2 (red squares) and $<$-0.2 (blue circles) plotted with the Gilli et al. (solid) and Treister et al. (dashed) models with two different column density ranges $>$10$^{22}$ cm$^{-2}$ in red and $<$10$^{22}$ cm$^{-2}$ in blue. The ratio of \leg\ number counts to Gilli et al. (2007; solid) and Treister et al. (2009; empty) models are plotted in the soft and hard bands (bottom left and bottom right).}
\label{hr_numbercounts}
\end{figure*}

The Gilli et al. (2007) and Treister et al. (2009) models are based on different assumptions on the fraction of obscured sources and on the assumed luminosity and redshift dependences. Therefore, their differences are more marked when considering obscured and unobscured sources separately. We used the hardness ratio, as defined in Section \ref{hr}, to divide the sample using HR$>$-0.2 for obscured sources and HR$<$-0.2 for unobscured sources. In the soft (hard) band there are 1057 (1332) obscured sources and 1701 (911) unobscured ones. 
In Figure \ref{hr_numbercounts}, we present the number counts in the soft and hard bands (left and right) for both obscured (red) and unobscured (blue) sources. 
A clear difference is observed in the number counts of obscured and unobscured in the soft band, where we observe a ratio of up to $\sim$10 at bright fluxes, while it almost disappears in the hard band, where the ratio is very small at all fluxes. This implies that the difference must be dictated by obscuration effects. 

The models from Gilli et al. (2007, solid line) and Treister et al. (2009, dashed line) assuming column densities above and below 10$^{22}$ cm$^{-2}$ (red and blue respectively) are plotted in the same Figure. In the soft band, both predictions of the number of unobscured sources are in agreement within 5\% with our data, up to fluxes of $\sim$3$\times$ 10$^{-14}$ \cgs, while the difference becomes larger for obscured sources ($>$10--20\%), with both models over predicting the number of sources at all fluxes. In this last case, the Treister et al. (2009) model predictions are generally worse than those of the Gilli et al. (2007) model, by 5--10\%. In the hard band instead, model predictions are in general excellent agreement with our data (differences $<$5\% up to fluxes of 5$\times$10$^{-14}$ \cgs), for both samples above and below HR=--0.2. 

Overall, these discrepancies between data and models are totally expected given that as for example a different spectral model could change source fluxes and sky coverage, and that the spectral parameters in the Gilli et al. and Treister et al. models are different from those used in this work. Therefore, despite all the underlying assumptions, the differences between observed number counts and phenomenological models are remarkably small ( 2-5\%; see also LaMassa et al. 2013a for a discussion on discrepancies between data and population synthesis models).

\section{Summary and Conclusions}
\label{summary}

In this paper we presented \leg, a 2.2 deg$^2$ \chandra\ survey of the COSMOS field. We employed a total of 4.6 Ms of exposure time, including 1.8 Ms already published by E09 plus 2.8 Ms obtained as an X-ray Visionary Project during \chandra\ Cycle 14. The new data comprise 56 overlapping observations which, added to the 36 C-COSMOS pointings, yield a relatively uniform coverage of $\sim$160 ksec over the whole {\it Hubble}-covered area. By construction, the survey flux limit is the same of C-COSMOS, computed in three bands using the same approach of P09.

We followed the same procedure used and tested by P09 combining standard CIAO tools for the data reduction, and \texttt{PWDetect} and \texttt{CMLDetect} for the data analysis, including the source detection and photometry. We also performed aperture photometry for consistency with the E09 and P09 analysis. The analysis was performed on the new \chandra\ data and also on the outer C-COSMOS frame, overlapping with the new observations. Given that the survey properties (exposure, roll angle and background counts) are consistent with C-COSMOS ones, we used the same probability threshold for the source detection corresponding to DET\_ML =10.8. At this limit, we detected 2273 sources that were not previously detected in C-COSMOS, by combining detections in the full, soft and hard bands. 385 of these sources were detected in the area overlapping with C-COSMOS: in the same area we have also found 676 of the 694 old detections, while 18 sources were not detected again. The total number of sources in \leg\ is 4016. 
The source properties, including counts, count rates, fluxes in three bands (full, soft and hard) and hardness ratio computed using a Bayesian approach are reported in an online table published with this paper. 

We computed the source number counts in both the soft and hard bands and we find good agreement between our results and other surveys in the literature as listed above. The large number of sources in \leg\ (20\% or more than the sources in other contiguous surveys) allows to constrain the number of counts at medium fluxes ($\sim$10$^{-15}$ \cgs) with 10\% errors and to reduce the uncertainties on the normalization at bright fluxes where discrepancies between different surveys still exist. The combination of \leg\ with other surveys at fainter and brighter fluxes allows to cover more than 4 orders of magnitude in flux. 

Using the hardness ratio we measure a fraction of obscured sources of 50$^{+17}_{-16}$\%, defined as sources with HR$>$--0.2, corresponding to column density $>$ 10$^{22}$ cm$^{-2}$ at all redshifts,  despite the uncertainties on the classification due to complex spectral modeling not taken into account in this work (see Wilkes et al. 2009, 2013). For the first time, we computed the number counts for obscured and unobscured sources separately using the hardness ratio as an indication for obscuration (HR=--0.2 corresponding to the separation between $>$ and $<$ 10$^{22}$ cm$^{-2}$). 
The large number of sources in each sample (about a thousand or more) allows to compute the number counts for the two populations and reveals a larger difference (in both normalization and shape) in the soft band, while a very small if not absent difference in the hard band is observed (the normalization is consistent, while we can observe a small difference in shape). Given the large range of luminosities and redshifts probed by \leg\, this can be interpreted as a difference in orientation rather than an intrinsic difference due to an evolutionary state between obscured and unobscured sources.

\begin{figure}[t]
\centering
\includegraphics[width=0.5\textwidth]{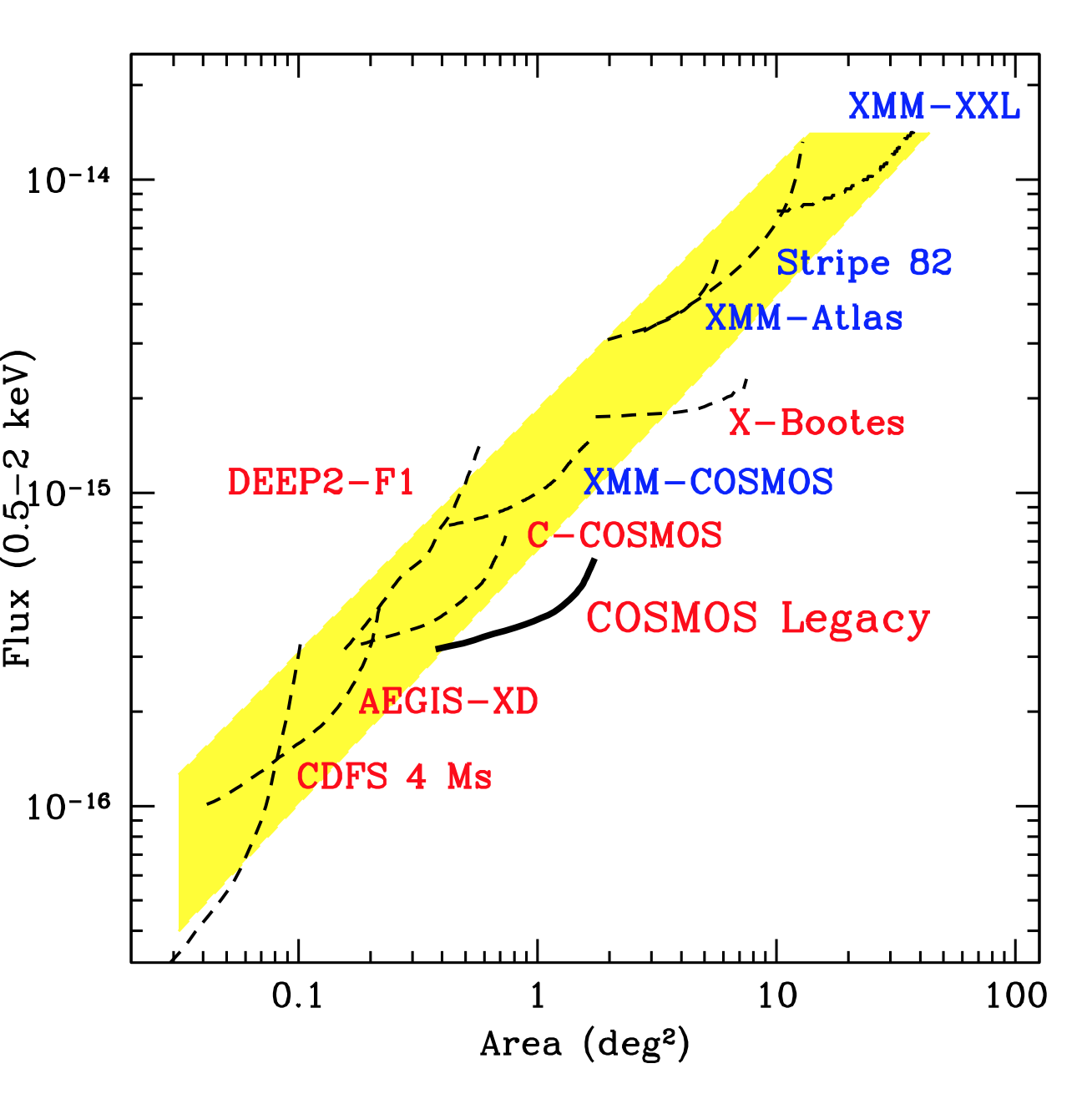}
\caption{Area--flux curves for \chandra\ (red) and \xmm\ (blue) contiguous X-ray surveys. Each survey has been plotted using each sensitivity curve starting from the flux corresponding to the area that
is 80\% of the maximum area for that survey to the flux corresponding to the 20\% of the total area. The plotted surveys are: CDFS 4Ms (Xue et al. 2011), XDEEP2-F1 (Goulding et al. 2011), AEGIS-XD
(Nandra et al. 2015), C-COSMOS (E09), XMM-COSMOS (Cappelluti et al. 2009), X-Bootes (Murray et al. 2005), XMM-Atlas (Ranalli et al. 2015), Stripe 82 (LaMassa et al. 2013a,b, 2015), XMM-XXL (PI: Pierre; see also Pierre et al. 2004). The survey locus described in the last section is drawn in yellow.}
\label{surveys}
\end{figure}

In Figure \ref{surveys}, the area -- flux parameter space of the most recent \chandra\ and \xmm\ surveys (CDFS 4Ms, Xue et al. 2011; AEGIS-XD, Nandra et al. 2015; XDEEP2-F1, Goulding et al. 2012; C-COSMOS, E09; XMM-COSMOS, Cappelluti et al. 2009; X-Bootes, Murray et al. 2005; XMM-Atlas, Ranalli et al. 2015; Stripe 82, LaMassa et al. 2013a,b and LaMassa et al. 2015; XMM-XXL, PI: Pierre, see also Pierre et al. 2004) is presented. Most surveys lie on a locus (yellow shaded area) determined by our current X-ray telescope capabilities.
\leg\ is exploring a new region off this locus, an additional factor 2-3 deeper at the areas it covers, by using a total exposure time which is unusually large (4.6 Ms total) for that given area flux combination, and preparing for surveys with future facilities. The X-Bootes survey also explores a region off the survey locus, but at brighter fluxes and over a larger area. 

In future decades, with facilities like {\it eROSITA} (Merloni et al. 2012), {\it Athena} (Nandra et al. 2013) and {\it X-ray Surveyor} (Vikhlinin et al. 2012), it will be possible to explore a new region of area-flux parameter space, moving away from the current survey locus towards the bottom right corner of Figure \ref{surveys}. For example, {\it Athena+} will perform a multi-tiered survey and given the combination of large effective area and field of view will enable X-ray surveys to be carried out two orders of magnitude faster than \xmm\ and \chandra\ (see Figure~2 of Aird et al. 2013).  With a \chandra-like resolution over 10$^{\prime}$, {\it X-ray Surveyor} will be able to cover the same \leg\ area at the same flux in only 55 ksec (A. Vikhlinin private communication), 80 times faster than \chandra.

Thanks to the large area covered at considerable depth, \leg\ can now address those questions for which a large number of detected X-ray sources at a medium depth with uniform multiwavelength coverage and almost complete redshift information is needed. The excellent positional accuracy allows to obtain multiwavelength identifications and photometric redshifts for 96\% of the sources (Marchesi et al. in press). We are currently working on papers on the X-ray luminosity function with a focus on the high redshift universe (Marchesi et al. in prep.), the X-ray spectral analysis and X-ray variability of the bright sample with a focus on the hunt for obscured sources (Lanzuisi et al. in prep.), the multiwavelength spectral energy distribution fitting with host galaxy properties (mass and star formation rates) for both optically classified as obscured and unobscured sources (Suh et al. in prep.), clustering measurement and dark matter halo mass (Allevato et al. in prep.) and, finally, a catalog of X-ray extended sources (Finoguenov et al. in prep). 

The wide area and the availability of extensive multiwavelength data in the COSMOS field enable us to probe the average X-ray emission of objects not individually detected by \chandra, therefore  beyond the flux limit, through a stacking analysis. The combined \chandra\ \leg\ dataset is now fully implemented in the web-based \chandra\ stacking tool CSTACK\footnote{See http://lambic.astrosen.unam.mx/cstack. Login as user=guest, password=guest and see the explanatory manual. As of writing this paper, stacking analyses utilizing the C-COSMOS dataset is publicly available. Analyses involving the whole \chandra\ \leg\ dataset is still proprietary and will be public in due course.}. This enables us to investigate the X-ray properties of differently selected samples, such as optical selected galaxies (e.g. Mezcua et al. 2015, in press, finding indications of weak AGN activity in low mass non-elliptical galaxies), highly obscured AGN selected using both infrared or radio criteria, and early AGN populations at $z>$5.

\section{Acknowledgments}
The authors thank the referee for the useful comments. F.C. and M.C.U. gratefully thank Debra Fine for her support to women in science, the \chandra\ EPO team, in particular J. De Pasquale, for creating the true color X-ray image. This research has made use of data obtained from the \chandra\ Data Archive and software provided by the \chandra\ X-ray Center (CXC) in the CIAO application package.

This work was supported in part by NASA Chandra grant number GO7-8136A (F.C., S.M., V.A., M.E., H.S.); PRIN-INAF 2014 "Windy Black Holes combing galaxy evolution" (A.C. , M.B., G.L. and C.V.); the FP7 Career Integration Grant ``eEASy'': ``Supermassive black holes through cosmic time: from current surveys to eROSITA-Euclid Synergies" (CIG 321913; M.B. and G.L.); UNAM-DGAPA Grant PAPIIT IN104113 and  CONACyT Grant Cient\'ifica B\'asica \#179662 (T.M.); Collaborative Research Council 956, sub-project A1, funded by the Deutsche Forschungsgemeinschaft (A.K.); NASA contract NAS8-03060 (T.A., A.F., K. G); the Greek General Secretariat of Research and Technology in the framework of the programme Support of Postdoctoral Researchers (P.R.); NASA award NNX15AE61G (R.G.); the Science and Technology Facilities Council through grant code ST/I001573/1 (D.M.A.); the Swiss National Science Foundation Grant PP00P2\_138979/1 (K.S.); the Center of Excellence in Astrophysics and Associated Technologies (PFB 06), by the FONDECYT regular grant 1120061 and by the CONICYT Anillo project ACT1101 (E.T.); 
the European Union's Seventh Framework programme under grant agreements 337595 (ERC Starting Grant, ``CoSMass'') and 333654 (CIG, AGN feedback; V.S.)
S.T. is part of The Dark Cosmology Centre, funded by the Danish National Research Foundation. B.T. is a Zwicky Fellow.


\end{document}